\documentclass[prd,aps]{revtex4}
\usepackage{amssymb,epsfig}

\begin{document}

\title{AMR, stability and higher accuracy}

\author{Luis Lehner$^{1}$, Steven~L.~Liebling$^{2}$ and Oscar Reula$^3$}

\affiliation{$1$ Department of Physics and
Astronomy, Louisiana State University, 202 Nicholson Hall, Baton
Rouge, Louisiana 70803-4001, USA\\
$2$ Department of Physics, Long Island University -- C.W. Post Campus, 
    Brookville, New York 11548, USA\\
$3$ FaMAF, Universidad Nacional de Cordoba, Cordoba, Argentina 5000}

\begin{abstract}
Efforts to achieve better accuracy in numerical relativity have
so far focused either on implementing second order 
accurate adaptive mesh refinement or on defining higher order accurate differences
and update schemes. Here, we
argue for the combination, that is a higher order accurate adaptive scheme. This combines
the power that adaptive gridding techniques provide to resolve fine scales (in
addition to a more efficient use of resources) together with the higher accuracy furnished by
higher order schemes when the solution is adequately resolved.
To define a convenient higher order adaptive mesh refinement scheme, we discuss a few 
different modifications of the standard, second order accurate approach
of Berger and Oliger. Applying each of these methods to a simple model problem, we
find these options have unstable modes. However, a novel approach
to dealing with the grid boundaries introduced by the adaptivity appears stable and
quite promising for the use of high order operators within an adaptive framework.
\end{abstract}


\maketitle

\section{Introduction}
There is growing concern in the numerical relativity community about
the accuracy that can be achieved as robust implementations
of the Einstein equations become available. As has been discussed
a number of times, the simulation of strongly gravitating astrophysical
systems in 3D requires the stable and accurate handling of a large range of time and
length scales~\cite{lehnerreview,milleraccuracy}. 
In the context of finite difference schemes, 
increasing resolution by using more grid points in a uniform grid
quickly becomes prohibitively expensive, particularly with multiple spatial
dimensions. Two ways of achieving better accuracy include
(i) the use of higher order  methods for derivative approximations and time integration,
and (ii) the use of a grid hierarchy which adapts to provide resolution only where
and when needed, so called {\em adaptive mesh refinement} (AMR).
Both strategies have been adopted by the community largely following
two independent paths. On one
hand higher derivative operators are 
being introduced \cite{multipatch1,multipatch2,thornburg,zlochower,gioeldave} and on the
other AMR techniques within a second order accurate implementation have been adopted 
for over a decade \cite{ChoptuikAMR}. (For recent efforts implementing second order AMR/FMR see for
instance \cite{refsAMR}/\cite{refsFMR})

One key reason behind these independent developments lies in the complications encountered when
implementing AMR techniques, especially related to the way the
(artificial) interface  boundaries are treated. Special care must be taken
when dealing with the conditions at these boundaries, in particular how derivative operators
should be modified near them. In the case of second order accurate implementations, there is little
ambiguity on what these modifications can be as they need only be done at interface points. For higher
order implementations, the strategy is far less clear.
However, there are strong reasons to combine AMR and implementations of higher orders. For instance,
although higher order operators (be they within finite difference, finite elements
or spectral approximations)
do in principle provide an approximation with errors which are considerably smaller, they do so only
if the underlying solution involved is well resolved by the adopted grid. Certainly, the required grid can 
be anticipated  for linear problems but this is hardly the case in non-linear ones where finer features induced
by the non-linear nature of the problem can appear during the evolution. Hence, an adaptive gridding mechanism
which reacts to the solution itself is crucial. The combination of AMR with higher order operators then provides
lower errors for a given grid while being able to react to unanticipated features in the solution ---in addition to
providing a more efficient use of computational resources---.

In order to combine these techniques, one must face how to treat the derivative operators near
boundaries, and how to deal with the artificial boundaries themselves. A clean way to approach
this problem is to consider  operators satisfying summation by parts (SBP) which have recently been incorporated 
successfully in strongly/symmetric hyperbolic systems in numerical relativity~\cite{lsu1,lsu2,lsu3,multipatch1,multipatch2}.
These techniques guarantee that a large
set of problems can be implemented stably.

Having stability of the unigrid evolution,  however, is not sufficient to guarantee
a fully discrete energy estimate for an AMR implementation. The analysis of the time integration,
in particular how different grids communicate with each other to provide boundary conditions, becomes
quite involved. Nevertheless, one can argue strongly for adopting as a starting
point an implementation in which a semi-discrete estimate is guaranteed since a necessary condition for 
an AMR implementation to be stable is that its unigrid version be stable.   As discussed in the literature,
the latter is guaranteed when employing Runge Kutta operators of orders higher than 2 \cite{tadmor} (in
conjunction with suitable boundary conditions).

We thus adopt a strategy where spatial derivative operators satisfying SBP, along with Runge Kutta time integrators,
are employed on all levels of the AMR hierarchy. This choice, of course, does not
dictate the treatment of the interface boundaries, and we consider various options for this treatment.
For instance, in the standard Berger--Oliger strategy, the solution obtained at a parent grid is
employed to define boundary conditions for the child grid.  Leaving the stability issue aside for a moment,
an immediate observation is that when employing a two-level implementation, this strategy can not generically
give better accuracy than second order if boundary values are obtained by linear interpolation in time of 
the parent grid values. 
Additionally, the definition of all field values
via interpolation --without regard for their propagation features-- is a likely  source
of inconsistencies making stability a delicate issue.

In this work we examine alternative treatments of grid boundaries and derivative operators to deal with these issues. We pay particular
attention to the stability of the system by effectively implementing a toy model problem
within a fixed two-grid hierarchy and analyzing the spectrum of the overall update
operator.  These studies demonstrate that implementations with accuracy orders greater than two, 
employing the standard  Berger--Oliger scheme have unstable modes. Furthermore, related schemes share this unfortunate feature. 
This instability can be remedied, with some caveats, with the addition of
dissipation though at the expense of larger errors in the approximation. 
Additionally, we observe that a simple modification of the
standard strategy yields a stable scheme without requiring the
addition of dissipation, albeit at additional computational cost. In this new strategy, a child grid effectively never  
requires boundary points from a parent. This can
be achieved if the solution at level $n+1$ is obtained by integrating using the
past domain of dependence of the child grid only.  At first sight this is more computationally 
demanding as it requires the area 
of the child grid at level $n$ to be larger than that at $n+1$ so that the intersection of the numerical domain of
dependence of the solution at the level $n+1$ is completely contained in the refined region at level $n$. 
This is achieved simply by defining an `extended' child grid at level $n$ and discarding 
a buffer zone at level $n+1$. As we will see later in Appendix~\ref{sec:cost}, this extra cost need not be relevant
at all, as for a given target error, a coarser base grid treated this way might be comparable to
a finer base grid treated with the standard Berger--Oliger scheme plus the addition of dissipation.

This work is divided as follows. In Section~\ref{sec:technique}, we detail different boundary treatments
and describe our method to assess their stability properties.
In Section~\ref{sec:tests}, we present one and three dimensional tests which illustrate the behavior
observed. Section~\ref{sec:conclusion} concludes with some final comments, and we defer to 
Appendix \ref{sec:cost} the discussion of the cost associated with 2nd vs. 4th order implementations.

\section{Technique and stability issues}
\label{sec:technique}
In searching for a stable AMR implementation, a basic requirement is
to have a stable implementation of the equations in the uniform grid case.
This requirement is perhaps the most important building block, without which an AMR
implementation will suffer even more in terms of stability.
Unfortunately, even when ensuring the unigrid scheme is stable, no results are available
on what guarantees the stability of an AMR/FMR implementation with `sub-cycling' for arbitrary
accuracy orders\footnote{For
partial related results see\cite{berger,trompert,oligerzhu}}.
Such a scheme which advances a grid with a time step defined by its grid spacing
stands in contrast to one which advances all points by a global time step determined by
the finest resolution in use.
The latter is considerably
less efficient than the former (becoming more so with more refinement levels) but
is simpler to analyze. We concentrate here on the former case, i.e. one that involves sub-cycling,
as this is the method one would like to employ.

In order to develop a stable AMR/FMR scheme, we analyze different natural options
from their fully-discrete point of view and study their stability properties.
We begin by adopting a strategy that ensures stability for a uniform grid case
for first order (non-constant coefficient) symmetric hyperbolic linear systems. This
strategy calls for employing derivative operators satisfying ``summation by parts'' (SBP)
together with an appropriate time integrator such as 3rd/4th order Runge--Kutta schemes\cite{tadmor}.
The first ingredient allows one to derive a semidiscrete energy estimate while
the second one guarantees this estimate will be preserved in the fully discrete
case\footnote{For more details the reader is referred to \cite{gko,olson,tadmor} and
for a presentation, extension and tests of these techniques within numerical 
relativity to \cite{lsu1,lsu2,lsu3}}. 
Following this strategy, we employ such operators here as our main building
blocks; as mentioned above, they
provide a strong platform for developing a useful AMR/FMR prescription.

Having defined the constituents of the basic update routine, we now turn our attention to
the creation and update of a child grid and how updated field values 
on this grid are to affect those in the parent one. Our goal is to obtain a recipe that will
allow us to distinguish among different possibilities, the ones that guarantee stability 
for a simple model problem. As we will show, even when applied to 
a simple model problem, the analysis we present does well at predicting either stability or 
instability, suggesting applicability for more general problems. Although our analysis applies
to a simple model problem, the scheme it reveals to be stable is one that will naturally yield
a stable scheme if it is so for a unigrid implementation. Since the adoption of operators
satisfying SBP together with an appropriate Runge--Kutta update scheme guarantees stability
for generic non-constant coefficients first order linear systems\cite{lsu1,lsu2}, 
the higher order AMR scheme will preserve this property.

To fix ideas and to simplify the presentation of our analysis, 
we restrict to the case where second and fourth order
accurate schemes are employed. This is sufficient to illustrate the salient aspects of
the analysis and main results; the conclusion can be trivially extended to arbitrary
orders of accuracy.

\subsection*{Analytical Considerations}
\label{sec:analytical}

In order to explore different possible ways to implement an AMR strategy 
we analyze what a full discrete step involves for the simple equation
\begin{equation}
\partial_t u = \partial_x u,
\end{equation}
on the region $x \in [0,1]$ with periodic boundary conditions, and where a refined region (the child grid) is included
within this domain. This can be assumed without loss of generality as the operators employed (satisfying SBP) and a suitable
Runge Kutta time update (together with suitable boundary conditions) guarantee the stability of the implementation. Thus, to 
avoid unnecessary clutter in the analysis we examine the periodic case for the coarsest grid involved.
For simplicity, we restrict ourselves to the case of {\it fixed mesh refinement} with
the child grid $2:1$ refined with respect to the parent grid.
We then compute the eigenvalues of the full amplification matrix ($Q$) corresponding to the update scheme of field values on the parent level. 
%
%
Clearly, if all the absolute values of this matrix are bounded by unity, amplification at each step is 
bounded by one (i.e. modes are either kept constant or damped) and stability follows for the problem under consideration
if the matrix can be diagonalized. 
On the other hand, if there is an eigenvalue with absolute value larger than unity, the
amplification per step is unbounded and the algorithm is therefore unstable. Since this applies already to the simplest
first order problem, the conclusion that the system is unstable will naturally hold for generic ones.
Note that it is sufficient to consider a single refinement level (and the corresponding coarse
grid) in this analysis, as in a generic case consisting of several child grids the total amplification 
matrix will be a composition of several $Q$ matrices.

The specific form of $Q$ depends on several different ingredients. The cleanest
way to construct $Q$ is to express it as a composition of different matrices, each representing
the action of an operator performing a specific task involved in the overall update strategy. 
For the problem under consideration, we need to consider the matrices
\begin{itemize}
\item $J_{21}$ which defines the grid values on the child level by knowing those at the coarse level, usually
referred to as prolongation or interpolation.
\item $D_{1}$ and $D_{2}$ the derivative operators used on the parent and child grids respectively.
\item $U$, the specific time stepping algorithm.
\item $J_{12}$ which restricts the values obtained on the child grid at the new coarse time level
at appropriate points of the coarse grid; usually referred to as restriction or injection.
\item $R$ a restriction operator which, when acting on an overall update scheme, restricts its values to the
region in the coarse grid which needs no refinement. Hence, $1-R$ restricts the action of any operator to the points
in the coarse level which do require refinement.
\item $B$, boundary points definition at intermediate child grid time levels which, depending on  the
strategy adopted, one might have to consider.
\end{itemize}
An update step in a two-level scheme which takes values on the $n$-th level  of the coarse
grid ($u_c^n$) and yields values at the $n+1$-th level ($u_c^{n+1}$) can be expressed as
$u_c^{n+1} = Q_c u_c^{n}$, with:
\begin{eqnarray}
Q_c \equiv  R U_c  + (1-R) J_{12} \,  U_f \, J_{21} ,
\end{eqnarray}
with $U_c$ and $U_f$ the update steps on the parent and child grids. For instance,
for the advection equation above, and employing a 3rd order Runge Kutta scheme,
$U_c$ becomes (for a periodic boundary problem)
\begin{eqnarray}
U_c = 1 + \Delta t D_1 + \frac{(\Delta t)^2}{2} D_1^2 + \frac{(\Delta t)^3}{6} D_1^3.
\end{eqnarray}
The expression for $U_f$ is more delicate and depends crucially on the scheme adopted.
For instance, when employing $2nd$ order derivative operators, and the standard Berger-Oliger
strategy for treating interface boundaries, one schematically has
\begin{eqnarray}
T^n_1 &=& B \left ( 1 + \frac{\Delta t}{2} D_1 \right) + (1-B)\, I_{n+1/4} \, , \\
T^n_2 &=& B \left ( 1 + \frac{3 \Delta t}{4} D_1 T^n_1 \right) + (1-B)\,I_{n+3/8} \, , \\
T^n_3 &=& B \left ( 1 + \frac{\Delta t}{9} (2 D_1 + 3 D_1 T^n_1 + 4 D_1 T^n_2) \right) + (1-B)\,I_{n+1/2} \, ;\\
Q_f &=& T^{n+1/2}_3 T^n_3,
\end{eqnarray}
with $T_i$ denoting the intermediate steps of the Runge-Kutta algorithm,
$B$ the identity operator at the interior points of the child grid, but zero at interface
boundaries; $I_{m}$ the interpolation of field values at points on the coarse grid (at interface boundary points)
providing values at the $m$-th time level required by the time stepping algorithm.

The above is just an example of what a typical procedure entails as there certainly are many
alternatives for the injection/prolongation, update, derivatives operators and boundary definition that
will be part of the AMR/FMR strategy. A partial restriction is introduced when choosing the accuracy
order desired as this fixes the derivative operators at interior points (by adopting the standard
centered operators). However this still leaves considerable freedom related to the way derivatives are
to be defined at points at, and close to, the boundaries, in addition to the definition of the
interpolating operators $I$ to be used (if so desired) in defining interface boundary points, the
injection and prolongation operators, etc. The analytical investigation of every option is an involved
task. To reduce the number of possibilities we adopt some simplifying, yet natural and commonly used,
assumptions.  Throughout this work we concentrate on two-level schemes. We consider interpolating
operators acting only at interface boundary points and, at most, a few additional points at each side of
the refined grid. As much as possible we adopt the same basic integration step for all grids to ensure
that a recursive type algorithm can be easily employed. Finally we concentrate on vertex-centered
schemes and child grids having points lying on a coarse grid points\footnote{As we will see,
these grid requirements will not be relevant in the strategy found to be stable}.
Even with these simplifications, obtaining the
eigenvalues of the amplification matrix is a cumbersome task. For this reason, we perform our
investigations with the aid of a computer algebra package (Maple) to numerically calculate the
eigenvalues. 

To do this, we define a grid composed of $N_1$ points describing the coarse
level and where $M<N_1$ points will be refined (the child grid has $N_2=2M+1$ points). We construct the matrices
defined above, obtain the amplification matrix and numerically calculate the eigenvalues. 
Since we perform these calculations for a fixed number of grid points (though we have varied $N_1$ and $M$ over several
values), a negative result (all absolute values less or equal than unity) does not constitute a stability proof, 
but certainly a positive result, (at least one eigenvalue with larger than unity absolute value) 
shows that the proposed algorithm is unstable. We can thus substantially narrow the search for likely stable ones and the 
tests considered in Section~\ref{sec:tests} lend additional strength to our analysis. As we will see, our analysis indicates
a preferred option which, intuition suggests,  will be stable for any problem where the unigrid case is stable.

The construction of the amplification matrix  proceeds in the following way.
First a whole time step is taken in the coarser grid, so that we can use, if so desired, the values 
at the next time step to impose boundary conditions for the intermediate time steps taken
on the finer grid. Once the recipe to provide these boundary conditions is given and the
difference operator on the finer grids chosen, the algorithm is pretty much fixed. Other alternatives
include not providing boundary conditions at all but rather to taper off the child grid as subsequent
intermediate child time levels are obtained. Additionally, one might consider other
variants at the level of the interpolation order on $J_{21}$, a possible averaging at the injection level, $J_{12}$,
or at the coarser grid, etc. We analyzed several of these options and report on their results in 
the following subsections.

Finally, as mentioned, we limit ourselves to Runge-Kutta time integrators  of order larger than two as these 
are known to lead to stable schemes if the semi-discrete system is stable \cite{tadmor}.

\subsubsection{Second Order Schemes}
\label{sec:second}
Here we use a third order Runge-Kutta time integrator (RK3) as the second-order Runge-Kutta method
is not known to guarantee the fully-discrete stable scheme if the semi-discrete scheme is stable. 
The population of the finer grid, given by the operator $J_{21}$, 
is carried out using four-point stencil interpolation (having an $O(h^4)$ error) on the child 
grid points lying in between coarse
ones and using a direct copy  of field values where the grid points coincide (see appendix B for
the interpolator's specific form). We consider the following cases:

\begin{itemize}
\item \textbf{Berger Oliger: (\cite{berger2})}
Here we reproduce the ``standard'' Berger-Oliger scheme and check its stability when used with RK3.
Figure \ref{fig:1} illustrates the points involved in  this scheme. Here the values at the diamond points, 
needed for the intermediate RK steps of the finer grid (squares), are obtained via linear interpolation
in time
employing values at the top and bottom points of the coarse grid (circles).  The coarse grid points are
already known at the advanced time step because it takes a full time step before the fine grid is advanced.
The dotted lines represent the RK steps intermediate to a full time step of the child grid.
On the fine grid, only the square points are evolved, using centered difference operators, and
at the end of the cycle the values of the fine grid coincident with coarser points are {\it directly injected}
into the corresponding coarse grid points.

\vspace{1cm}
\begin{figure}[ht]
  \begin{center}
    \input{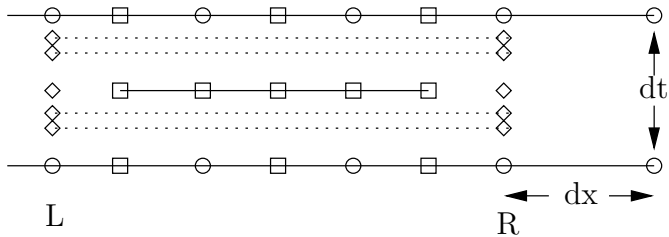}
    \caption{Standard Berger Oliger Scheme. Circles/squares denote coarse/fine grid points while
    diamonds denote boundary points at which values are filled via interpolation from the coarse grid. }
    \label{fig:1}
  \end{center}
\end{figure}

\item \textbf{Penalty Method:}
In this case, in contrast to the previous one, we pay close attention to the propagation
modes defined by the equation of motion. In this simple case, the single characteristic propagates
from right to left (see Figure \ref{fig:2}). Thus, only boundary values at the right end of the fine grid are required.
These are provided by linearly interpolating coarse field values at the top and bottom levels
and used as boundary conditions in a scheme enforcing these via penalty terms 
as described in \cite{carpenter0,carpenter1,multipatch1}. In this technique, boundary values are
incorporated by modifying slightly the right hand side of the (first order) equations. For our
test problem this results in
\begin{equation}
D u \rightarrow D u + P \kappa \frac{1+\delta}{dx} (u_b - u) \, ,
\end{equation}
where $D$ is the discrete derivative operator (satisfying SBP) employed;
$P=0$ for $i=1..N-1$ and $P=1$ at $i=N$ (i.e. only non-zero at the right interface boundary point);
$\delta$ is a free parameter and  $\kappa$ is related to the norm/derivative operator employed in
ensuring SBP is satisfied (see \cite{multipatch1} for a detailed description).

The penalty technique provides a clean and efficient
method to provide boundary values which, coupled with the use of operators satisfying SBP, guarantees
the stability of the implementation. Furthermore, within this approach, all points are updated and
the boundary conditions are imposed via a simple addition of suitable penalty terms to the right
hand side at boundary points. For these reasons it represents, at first sight,
a convenient approach to handle characteristic modes within an AMR/FMR implementation.
As with the Berger-Oliger scheme, after the child grid update to the same time as the parent,
field values coincident with the parent are directly copied.


\vspace{1cm}
\begin{figure}[ht]
  \begin{center}
    \input{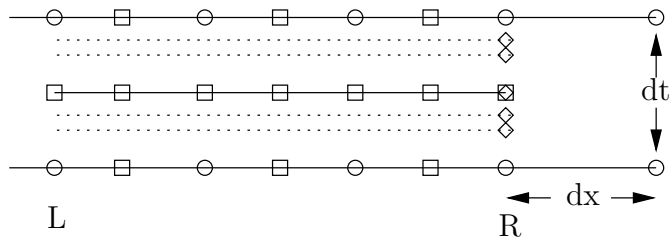}
    \caption{Penalty Scheme. Circles/squares denote coarse/fine grid points while
    diamonds denote boundary points at which values are filled via interpolation from the coarse grid.
    Notice only the points at the right are defined this way as the system describes modes propagating
    from right to left.}
    \label{fig:2}
  \end{center}
\end{figure}

\item \textbf{Tapered Boundary Method:}
In this case, we adopt restriction operators which effectively remove points that have been
influenced by the interface boundary. In practical terms, this means that if the
child grid at the advanced $n+1$ level is defined as $\{ x \in [L,R] \}$, the update from level $n$ must take
place in the region $\{ x \in [L-\triangle,R+\triangle] \}$ with $\triangle \equiv n_p dx$ given by a sufficient  number of
points to insure that the numerical domain of dependence of $[L,R]$ starting at $n+1$ is {\it completely}
contained in $[L-\triangle,R+\triangle]$. This is illustrated in Fig \ref{fig:3}. The size of $\triangle$ is
determined by the order of accuracy of the discrete operators involved and the maximum speed of the hyperbolic
problem considered.

Incidentally, note that this is not the same as an alternative approach
where ghost zones populating a region around the refined region (with values obtained
via linear interpolation from the coarse values at levels $n$ and $n+1$) are defined. 
That approach is relatively
straightforward to implement and reduces the complexity of the derivative operators employed, but
does not satisfy the requirement that what is done at the boundary has no influence on the
final injected point. As we will see later, this scheme also has unstable modes.

\vspace{1cm}
\begin{figure}[ht]
  \begin{center}
    \input{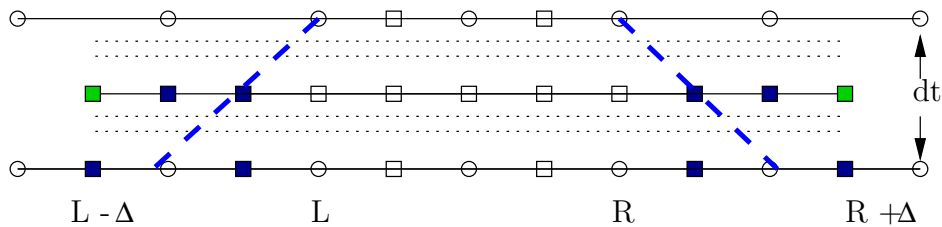}
    \caption{Tapered Scheme. Empty circles/squares denote coarse/fine grid points while
    filled squares schematically indicate those belonging to the fine grid which are affected
    by the presence of the boundary (either by derivatives at these being different from the centered
    one, and/or that their update reaches up to these points). Dashed lines indicate the
    past domain of dependence of the refined region at level $n+1$ and its intersection with level
    $n$ indicates the region that needs to have values in the fine grid. Hence, all fine grid points
    outside the domain of dependence are effectively discarded.}
    \label{fig:3}
  \end{center}
\end{figure}

\end{itemize}

\subsection*{2nd order schemes: Summary of results and observations}
\begin{figure}[ht]
\begin{center}
\includegraphics*[height=6cm,width=5.5cm]{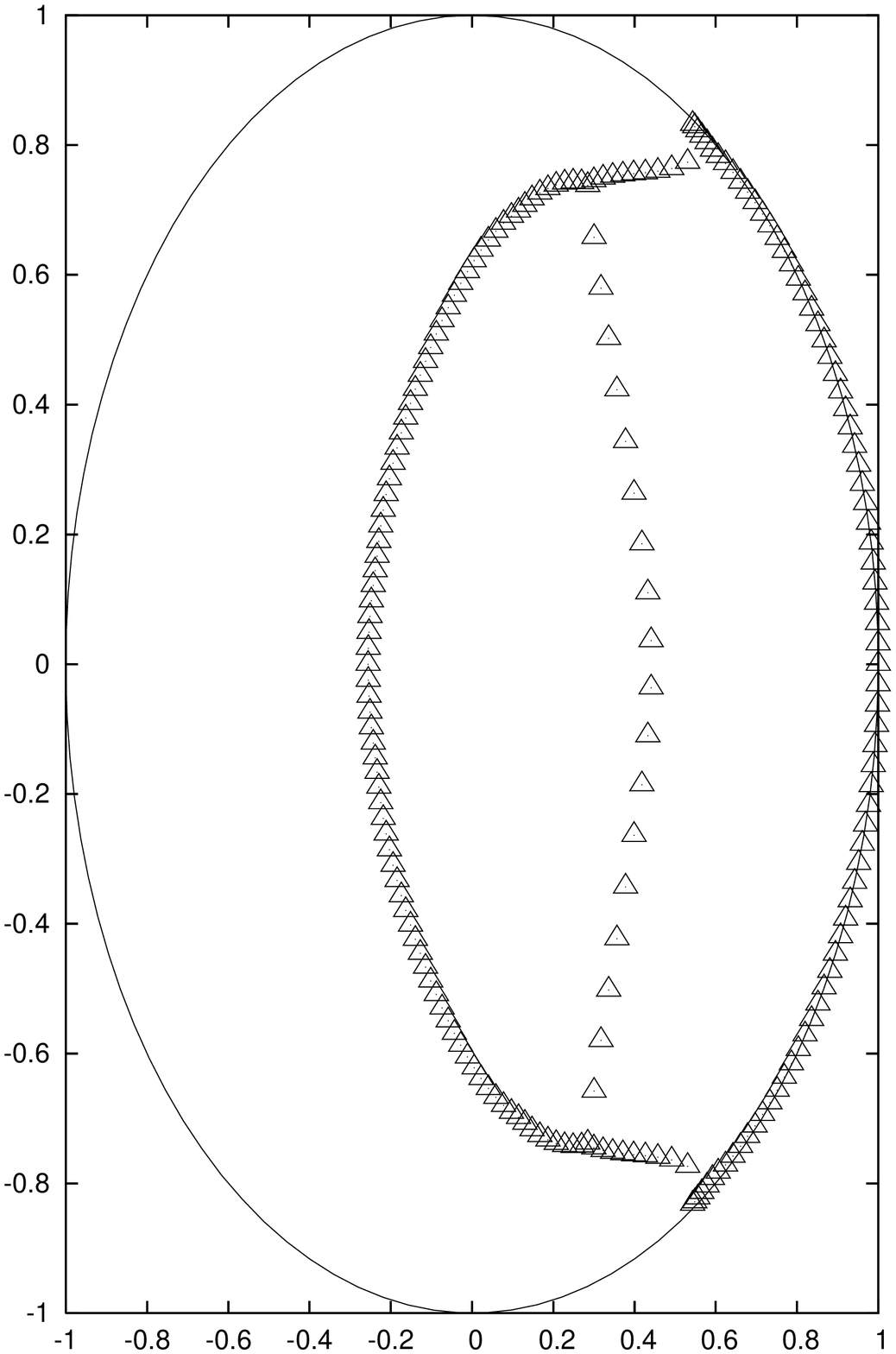}
\includegraphics*[height=6cm,width=5.5cm]{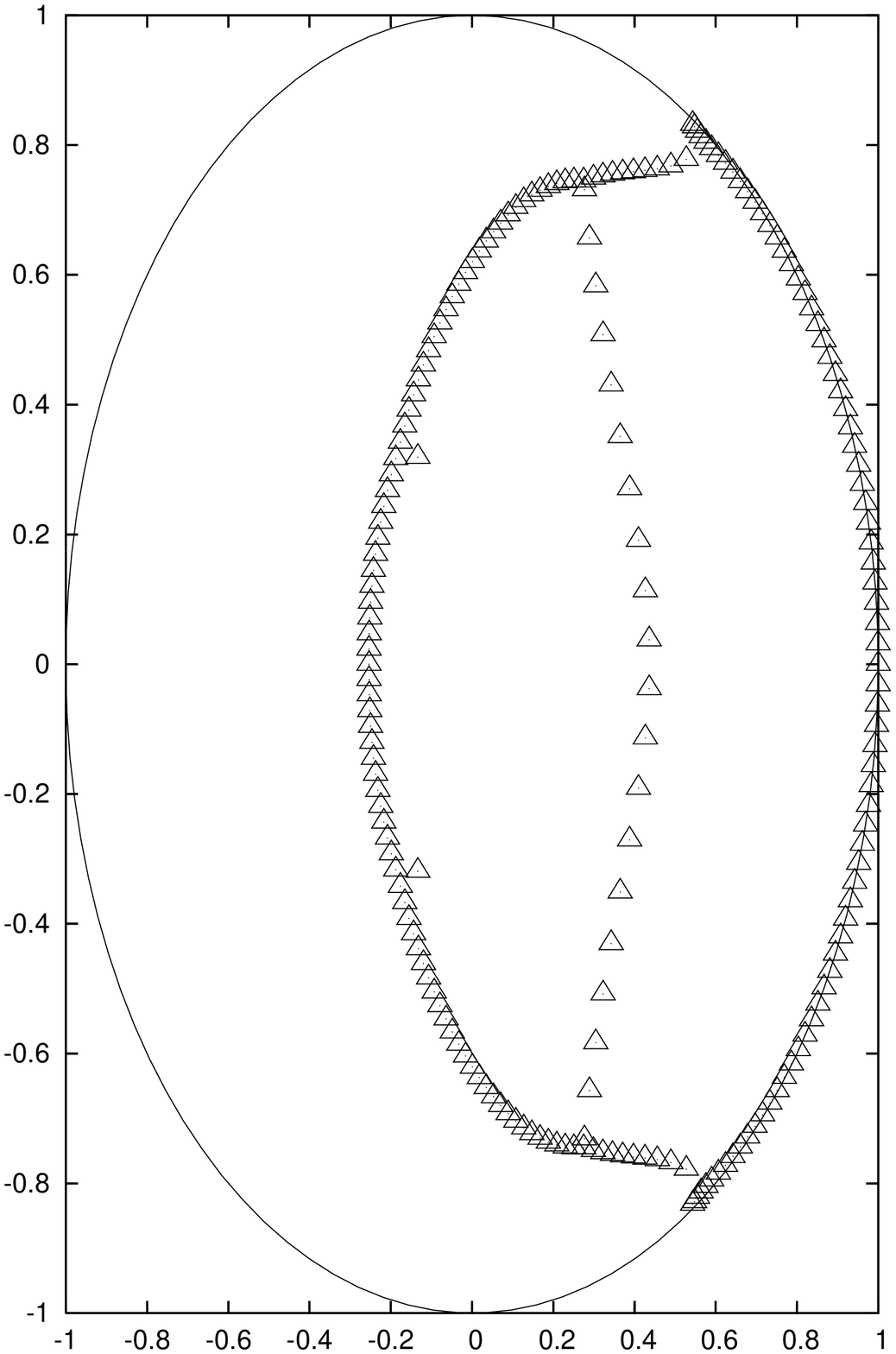}
\includegraphics*[height=6cm,width=5.5cm]{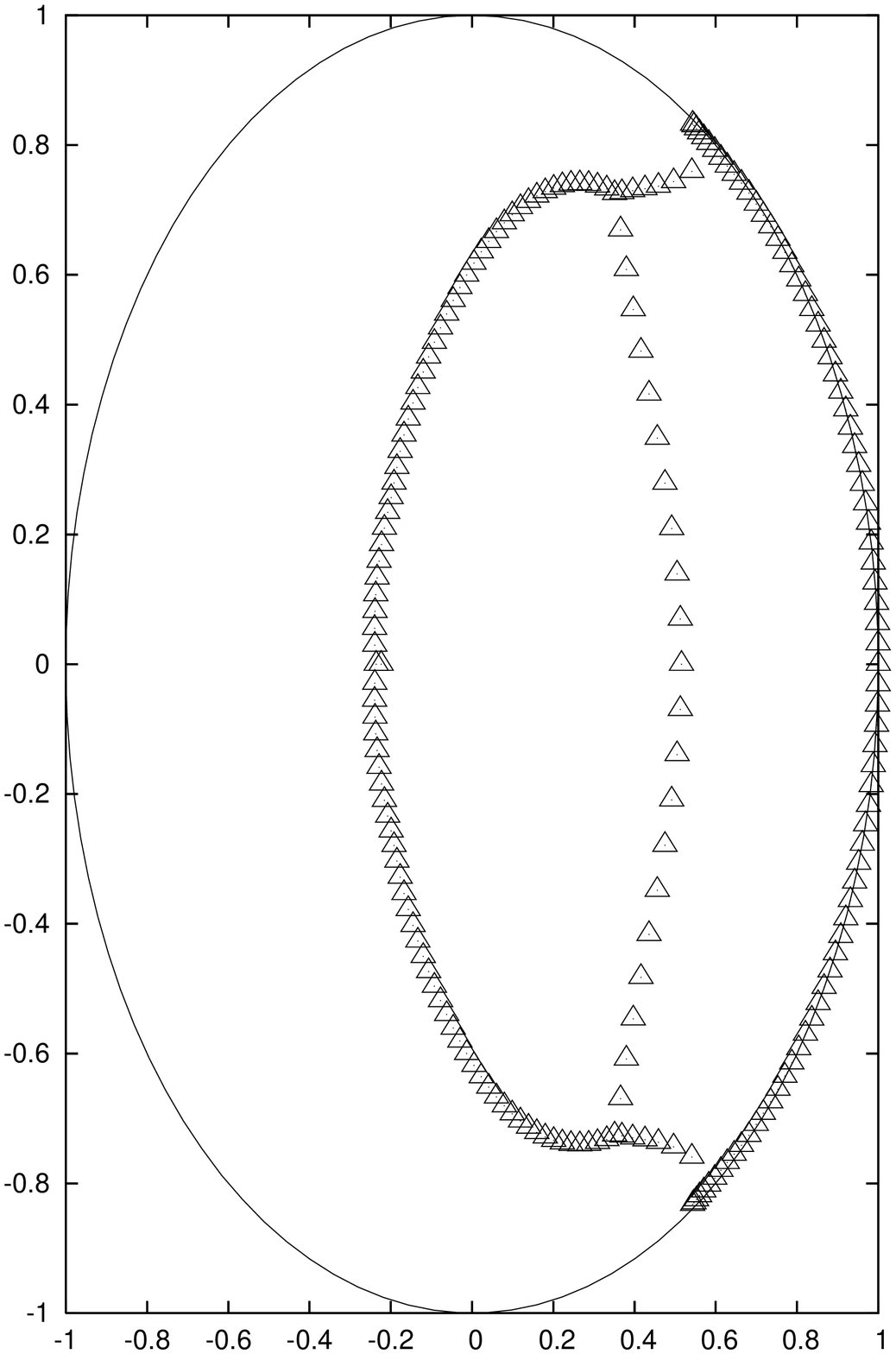}
\caption{Stability region of the fully discrete step for the Berger-Oliger, penalty
and tapered schemes (from left to right respectively).
For a CFL=1, all eigenvalues corresponding to each of the amplification matrices lie
within the unit circle.}
\label{BO2_eigenv}
\end{center}
\end{figure}

The analysis of the eigenvalues of the amplification matrices corresponding
to each of the three cases discussed above must be examined carefully as we are
considering the fully discrete case. Here a competition among the unstable modes
and the possible inherent dissipation of the update scheme determines whether
the eigenvalues' magnitude of the full amplification matrix are indeed larger than unity. 
In particular, for large CFL values, the inherent dissipation grows and the eigenvalues 
might be bounded by unity while for smaller CFL values this might not be the case. 
As we are interested in establishing stability with at most an upper bound for the CFL value
we have examined several possible values of CFL (ranging from $1$ to $10^{-4}$) and observe
when unstable modes arise.
For instance, for a CFL value of $1$ all eigenvalues for the three approaches lie
within the unit circle as illustrated in Fig.~\ref{BO2_eigenv}. However, as this value
is decreased unstable modes do appear for the penalty case while not for the other two options.
One can exploit the free parameter $\delta$ to alleviate the instabilities in the penalty case
(in the sense that the magnitude of the eigenvalues gets closer to unity) but not completely remove them
with the addition of artificial dissipation.

\begin{figure}[ht]
\begin{center}
\includegraphics*[height=7cm,width=5.5cm]{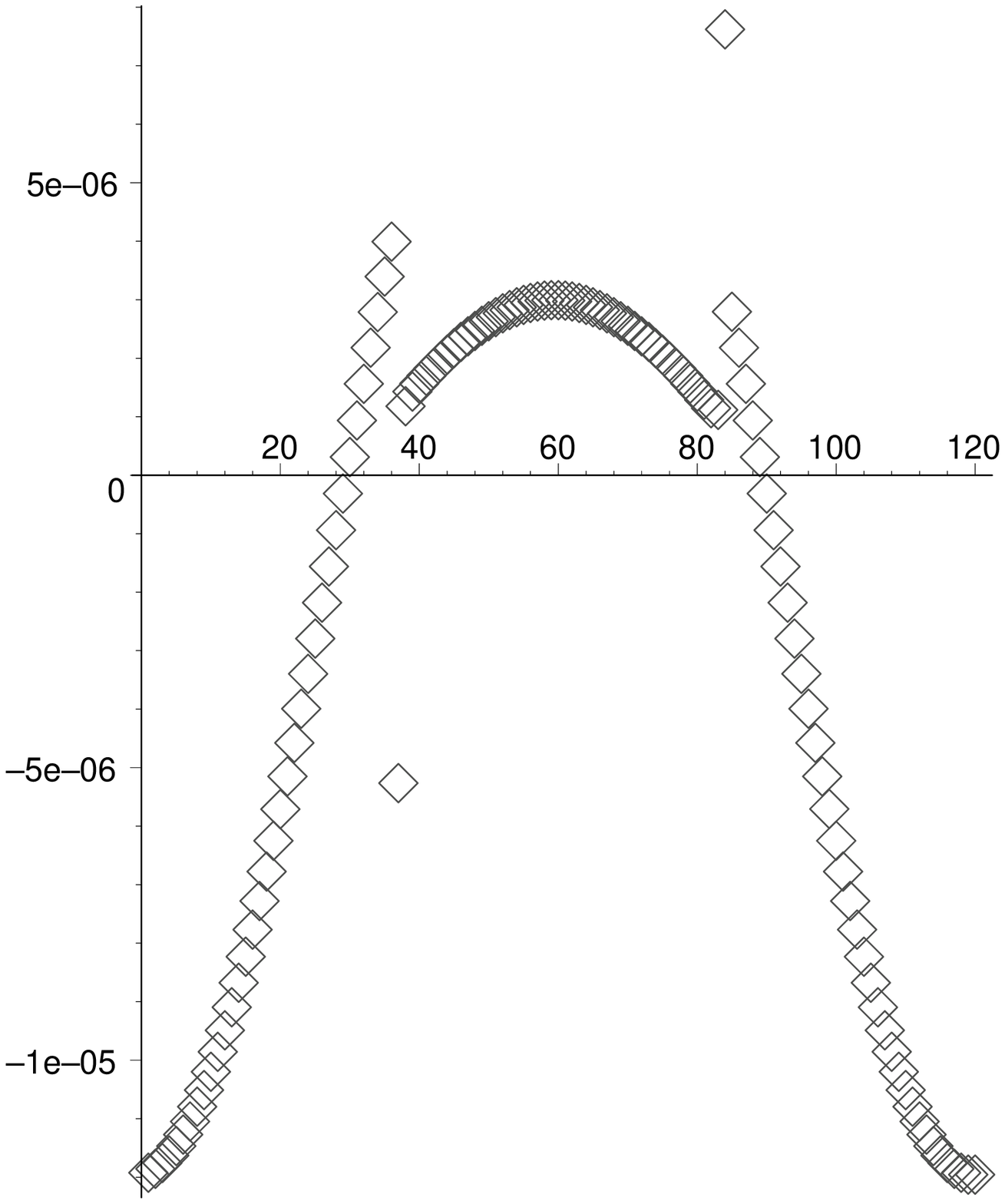}
\includegraphics*[height=7cm,width=5.5cm]{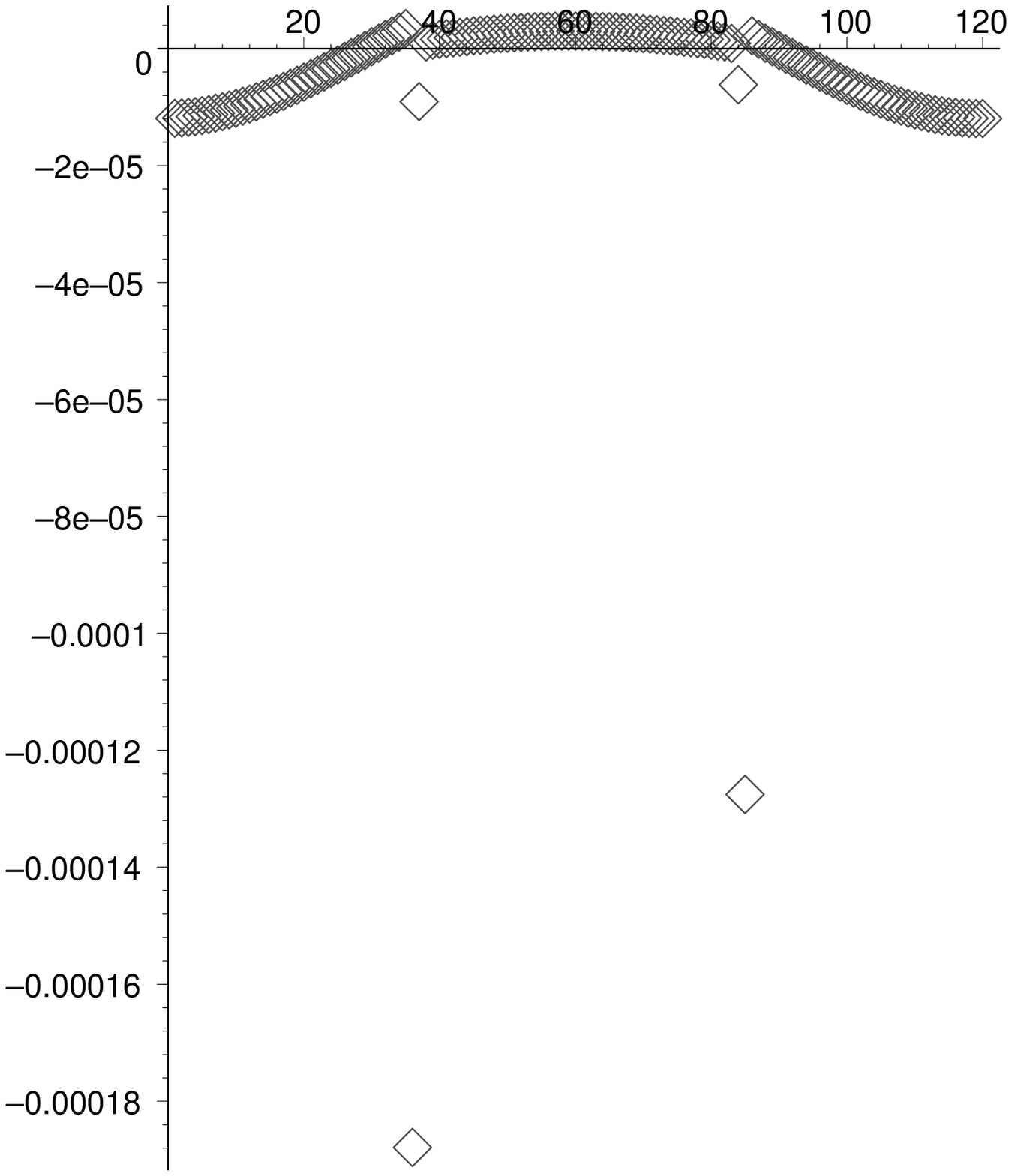}
\includegraphics*[height=7cm,width=5.5cm]{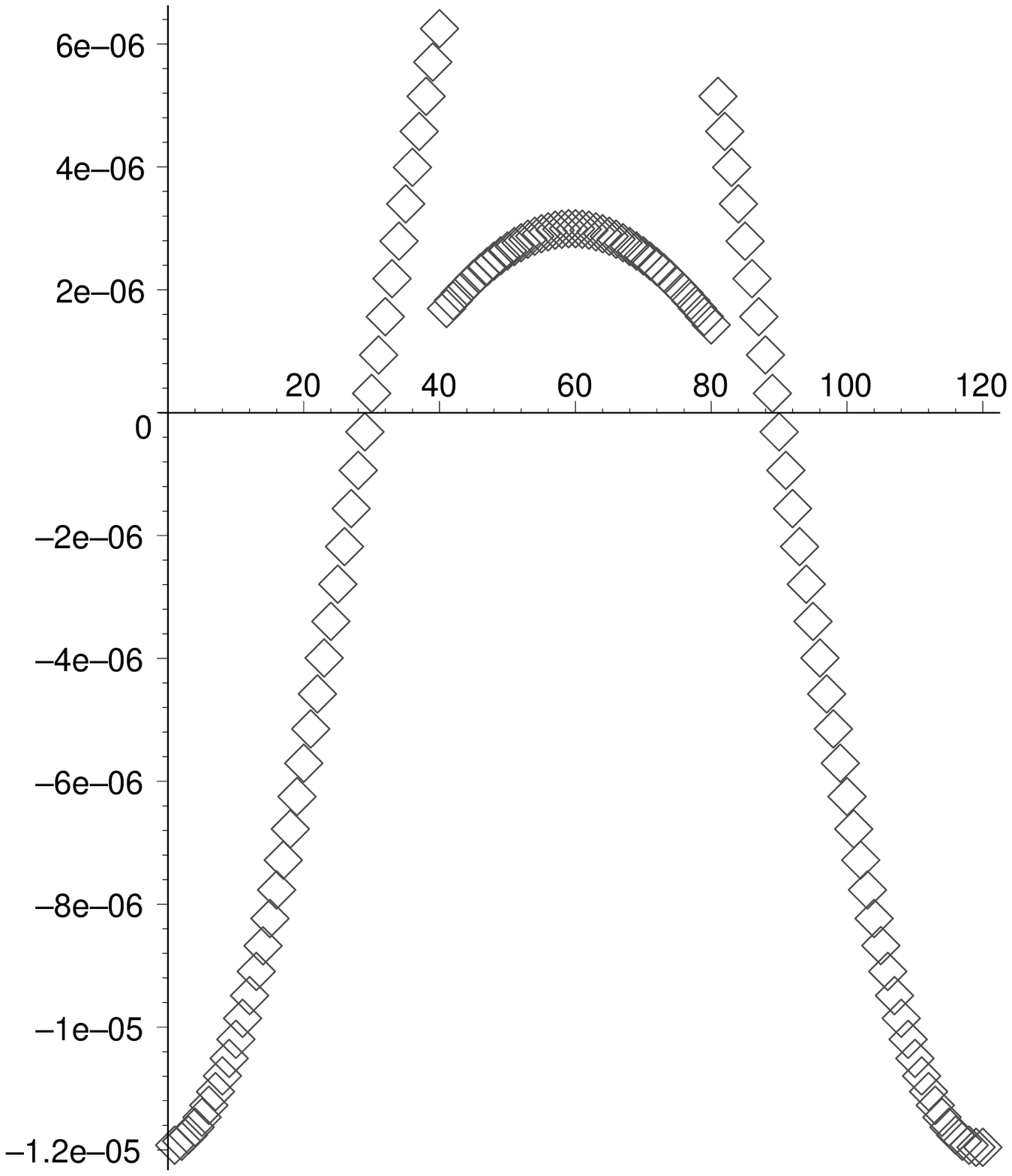}
\caption{Measured errors after one full step in the evolution of
$u=\sin(\pi x)$ in the domain $[0,1]$ under the three schemes considered 
---from left to right: the B-O, penalty and tapered schemes---. A CFL factor of
0.25 was employed and no artificial dissipation added.}
\label{O2_error}
\end{center}
\end{figure}

Additionally, it is instructive
to investigate the behavior of the error as a function of position to obtain some indication
of the effect the artificial boundaries have. To this end, we compute the error in the
evolution of $u(t=0,x) = \sin(\pi x)$ after a single step for each case. Here the solution 
after a step is $u(t=h,x) = \sin(\pi (x+h))$ and a point-wise error calculation is straightforward
to compute. The results are indicated in Fig.~\ref{O2_error}.
The differences in the errors among the three schemes are clear. The B-O and tapered approaches
have similar errors while those of the penalty scheme are larger
due to the derivatives
at the boundaries being calculated to first order accuracy.

\subsubsection{Fourth Order Schemes}
\label{sec:fourth}

Here we use a fourth order Runge-Kutta time integrator and analyze the stability properties
of the obtained schemes under the cases considered above. In this case, we will find that only
the tapered boundary approach yields a stable scheme unless a carefully tuned amount of
dissipation is added to the problem. We have found the same holds true 
when adopting a third order Runge-Kutta operator.
 
%
%
We adopt {\em fourth order accurate} centered difference operators (in the case of Berger-Oliger type schemes) or 
those satisfying the SBP property, in the case of the penalty approach. For the tapered boundary
approach it does not matter which one is employed as points affected by `special' boundary
treatments are to be discarded.
 
The population of child grid points lying between coarse ones uses a
six-point stencil interpolation operator (having an error of $O(h^6)$) to ensure that
the discontinuity in the truncation error in between different grids does not affect the order
of the scheme (see appendix B for the specific expressions of the interpolators). 
For a first order equation, employing fourth order derivative operators
requires at least a fifth order interpolation stencil. We adopt sixth order to avoid spoiling
the symmetry of the implementation.

We consider the following cases:

\begin{itemize}
\item \textbf{Berger-Oliger Method:}

Here the setting is as in the second order case, but some modifications must be introduced at the level
of the derivative operators employed. One option is to modify them near the boundaries in a side-ways manner
so as to preserve the order of the approximation. The other is to define extra ghost zones around the
refined region and keep the derivative operator in its standard centered expression. We have
analyzed both cases with similar qualitative results. We discuss here the latter case in some detail.

For a fourth order accurate approximation to the derivative operator only two ghost zones on each side of the 
grid are needed. One is located at the interface boundary while the other is placed next to it outside of
the refined region. As before, the field values at these points are to be defined via interpolation from
those known at the coarse level on levels $n$ and $n+1$. We examined both linear
interpolation --with error $O(h^2)$-- and quadratic interpolation --with error $O(h^3)$-- . The
latter option employs the value of the field at the $n$ and $n+1$ level (as in the linear case)
together with the time derivative of the fields at level $n$. The qualitative features of the amplification
matrix is the same in both cases, namely that there exists eigenvalues with magnitude greater than one.
We did not go beyond this order as it would involve taking further derivatives of the right hand side,
which in a general case would involve derivatives of both coefficients and variables

\item \textbf{Penalty Method:}
Here again the setting is similar to the second order one, the only difference is that at the boundary
the interpolation is done at third order, again using the computed right-hand-side of the equations at the initial
time. This is only needed at boundaries where there are incoming characteristics as at the other boundaries
field values will be defined solely by the update within the child grid itself. Additionally, no
extra ghost zones are required as one must use SBP operators of the appropriate order 
(see for instance, \cite{multipatch1}).

\item \textbf{Tapered Boundary Method:}
This case is analogous to the second order case. The only significant difference is the
size of $\Delta$, which must be enlarged so as to ensure the intersection of the past
domain of dependence of the region of interest at the $n+1$ level is completely contained inside
the refined region at level $n$.
\end{itemize}

\subsection*{4th order schemes: Summary of results and observations}
\begin{figure}[ht]
\begin{center}
\includegraphics*[height=7cm,width=5.5cm]{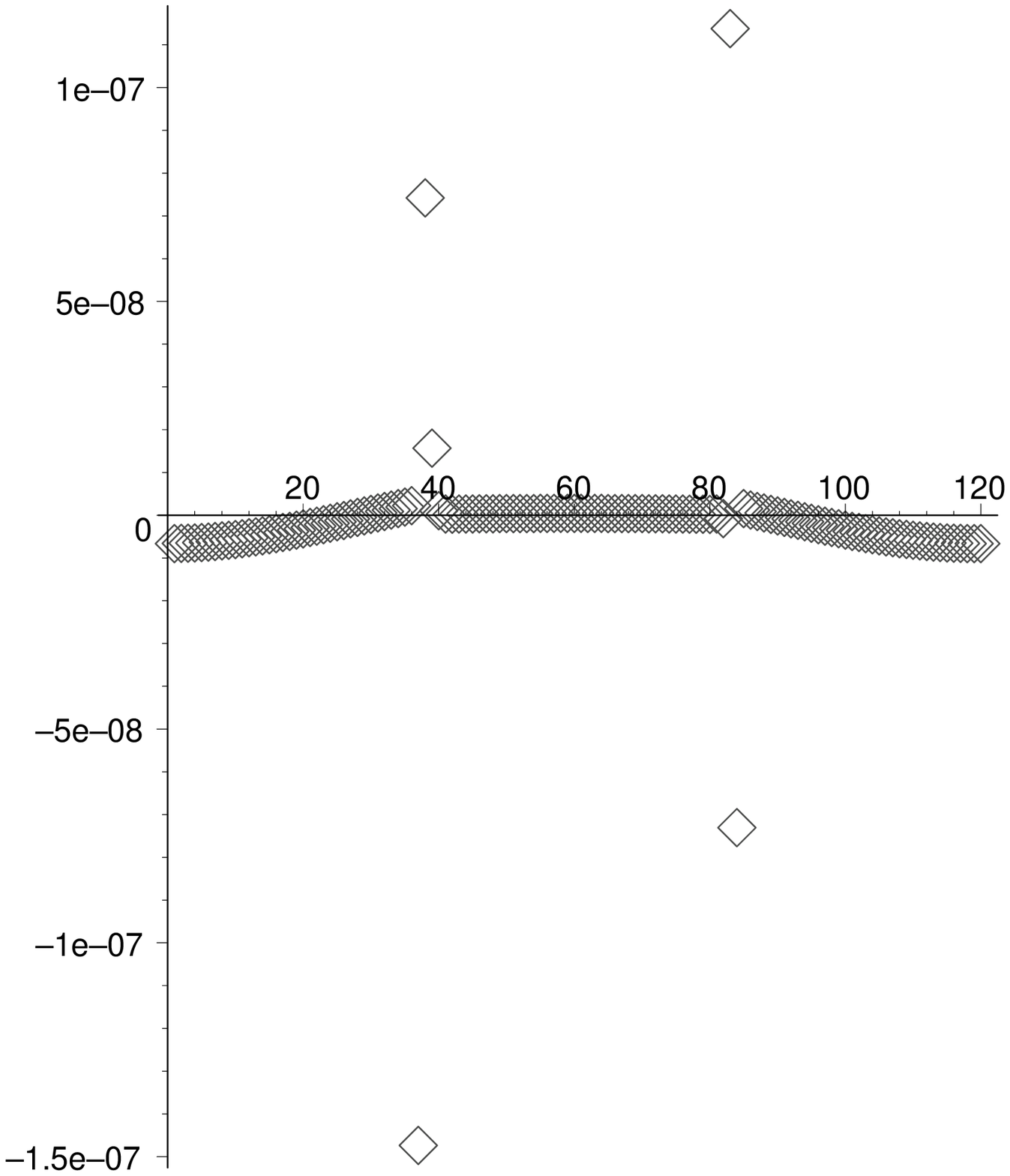}
\includegraphics*[height=7cm,width=5.5cm]{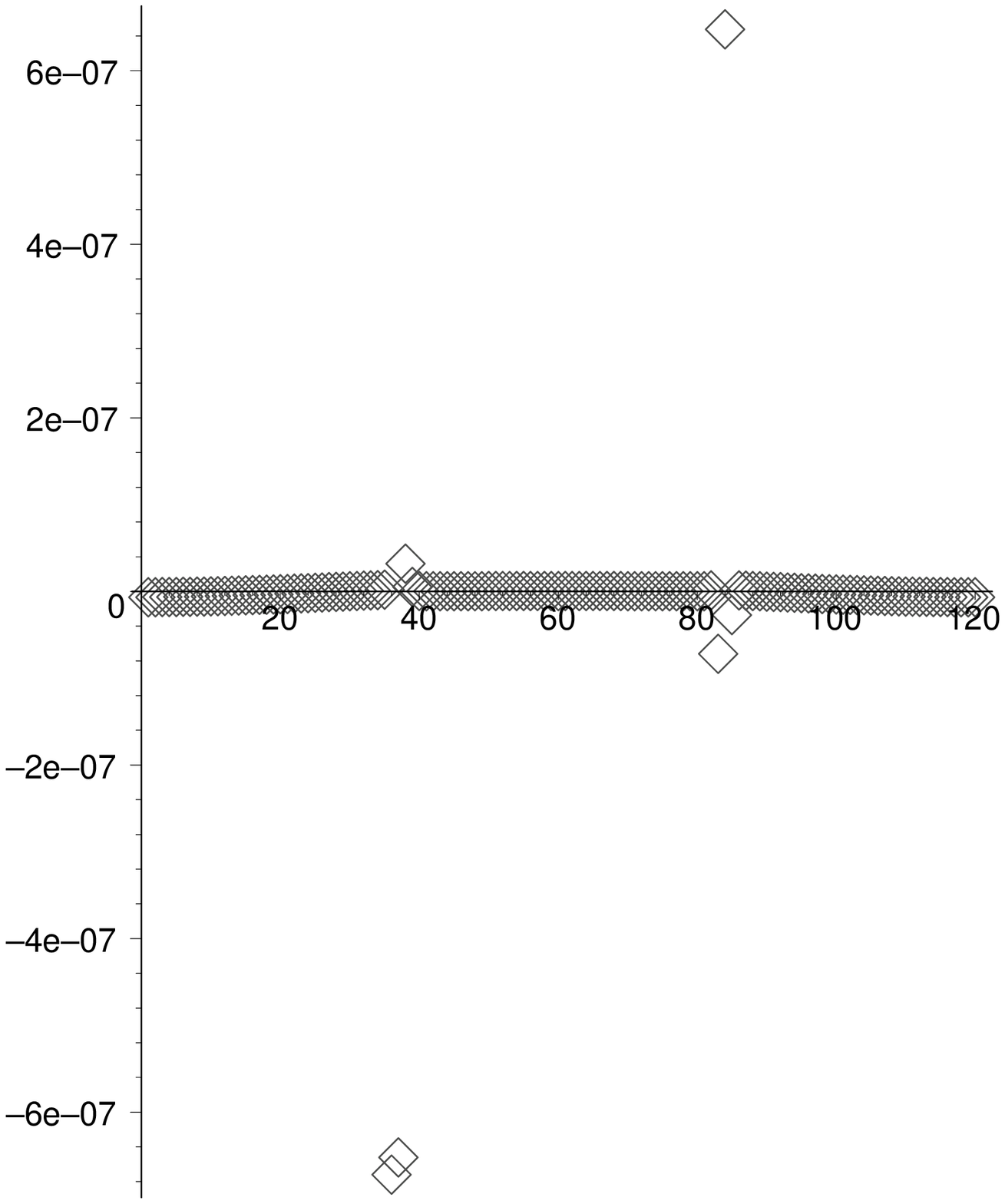}
\includegraphics*[height=7cm,width=5.5cm]{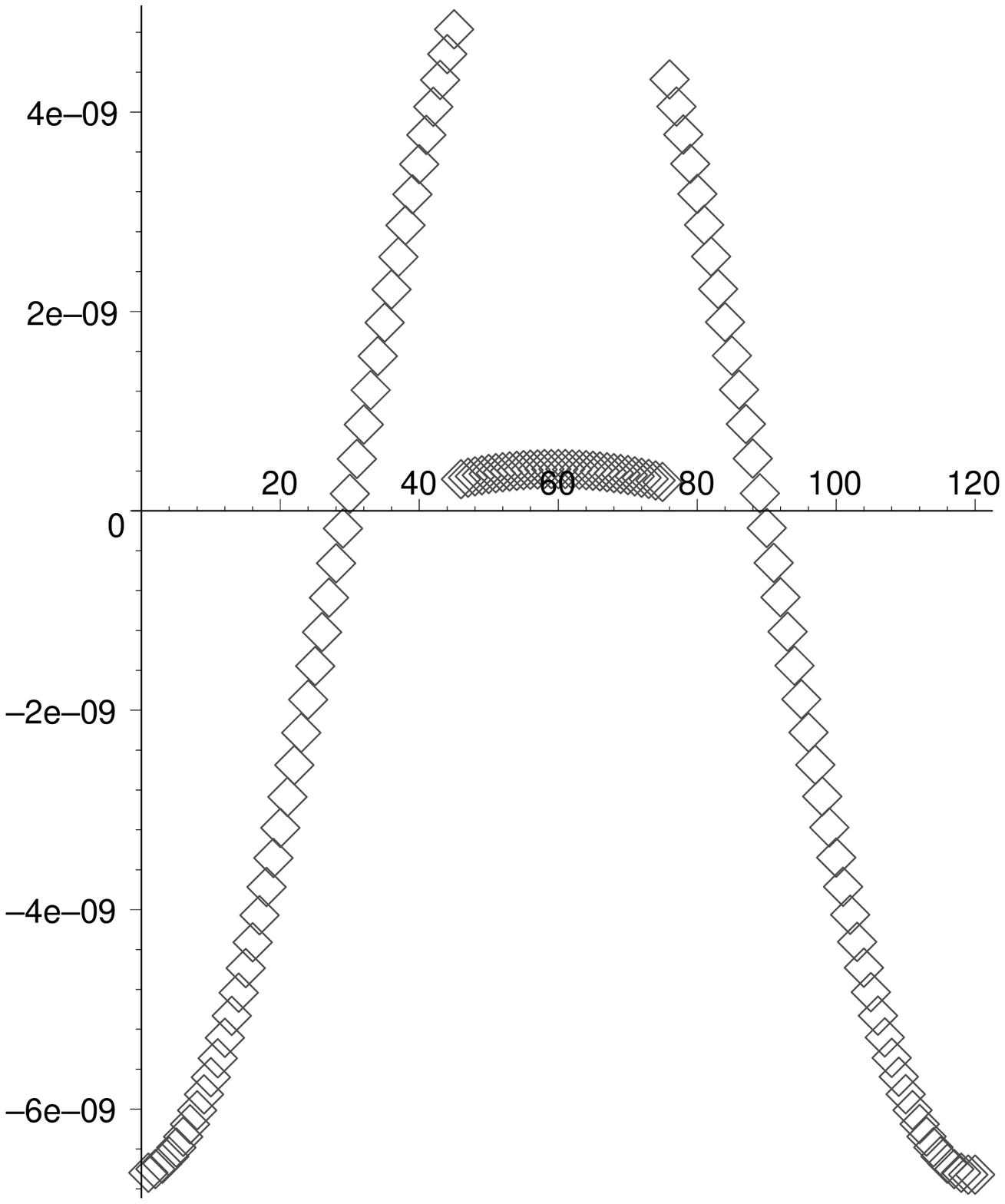}
\caption{Measured errors after one full step in the evolution of
$u=\sin(\pi x)$ in the domain $[0,1]$ under the three schemes considered 
---from left to right: the B-O, penalty and tapered schemes---. A CFL factor of
0.25 was employed and no artificial dissipation added.}
\label{O4_error}
\end{center}
\end{figure}

The analysis of the eigenvalues of the amplification matrices corresponding
to each of the three cases discussed above reveals that there exists eigenvalues with
magnitude larger than unity for the B-O and Penalty approaches while none for the tapered one. 
Clearly the stability of the tapered approach follows from the stability of the unigrid problem. 
Here again, the amount by which the eigenvalues of the B-O and Penalty approaches
are above unity is not too large and depends strongly on
the CFL factor employed. 
However, from our analysis it is expected that as the base grid is
refined and smaller CFL factors are used (or a larger number of refinement levels are employed) instabilities
associated with these modes will appear. Beyond this point however, the tapered approach also yields
evolutions with  much smaller errors. To illustrate this we again compute the error in the
evolution of $u(t=0,x) = \sin(\pi x)$. The results are indicated in figure \ref{O4_error}; again,
the effects of the boundaries can be
clearly appreciated but this time the errors in the tapered grid are 40 times smaller.

These results make it clear that straight forward extensions of Berger-Oliger to higher order, as well as 
the use of a penalty method which takes into account the characteristic structure of the problem, suffer
from unstable modes in a higher order AMR/FMR scheme.
The only option we have found to be free of this problem
is that given by the tapered boundary approach, which a priori would appear  more computationally demanding as points
are evolved which will be discarded. For sufficiently high order schemes, this can be a significant fraction
of the grid.

Although this extra cost might not be as significant as it appears (see Appendix~\ref{sec:cost}),
one might prefer to maintain the standard Berger-Oliger prescription and deal with the unstable modes
by introducing some amount of dissipation. Therefore we revisit the analysis above but now
with the addition of dissipation. To anticipate the results obtained, we indeed find
that dissipation can control the unstable modes but at the expense of increasing the overall error and
needing to fine tune it to ensure a desired degree of smoothness in the solution.

\subsubsection{Dissipation on Fourth Order Schemes}
\begin{figure}[ht]
\begin{center}
\includegraphics*[height=7cm,width=5.5cm]{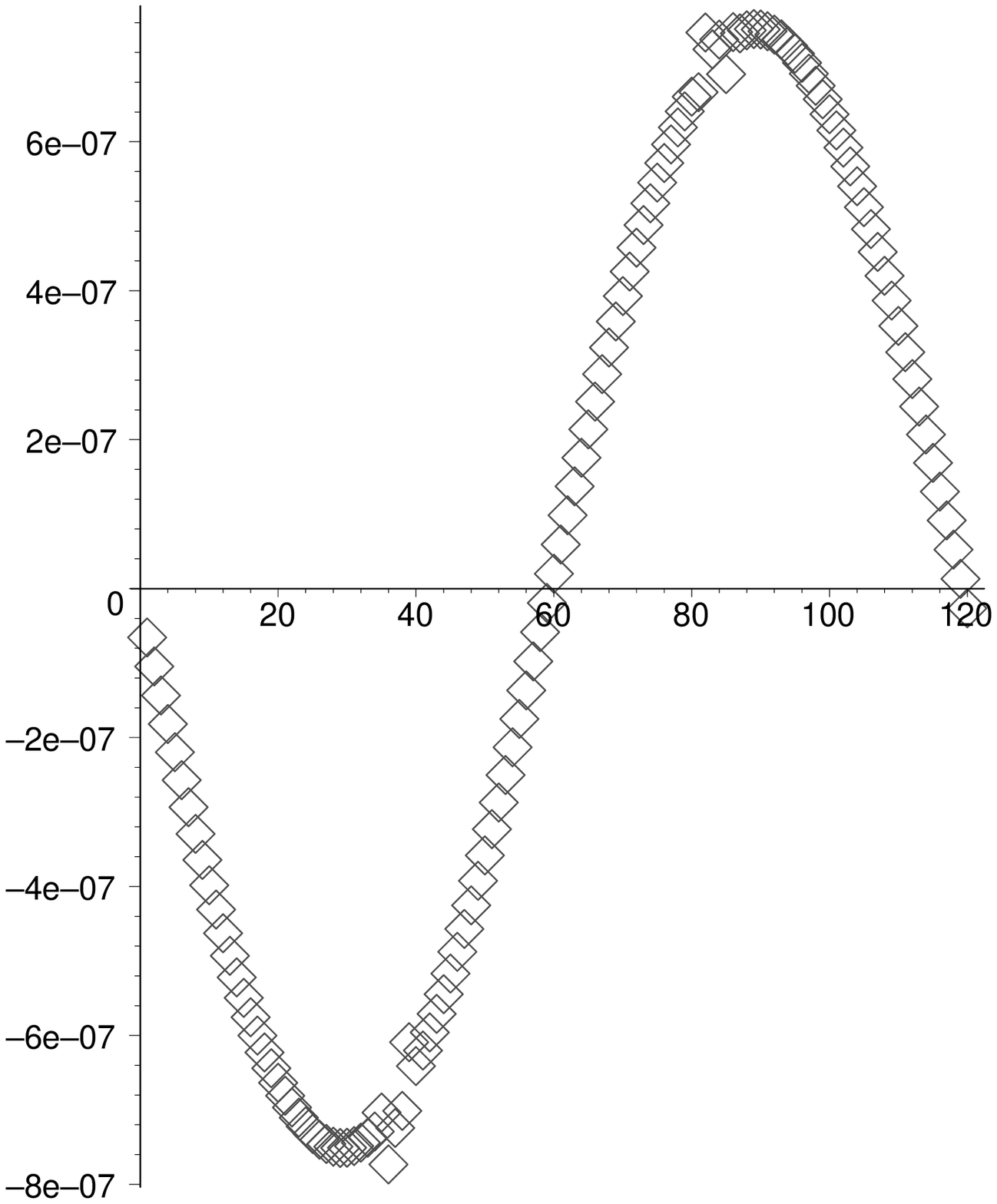}
\includegraphics*[height=7cm,width=5.5cm]{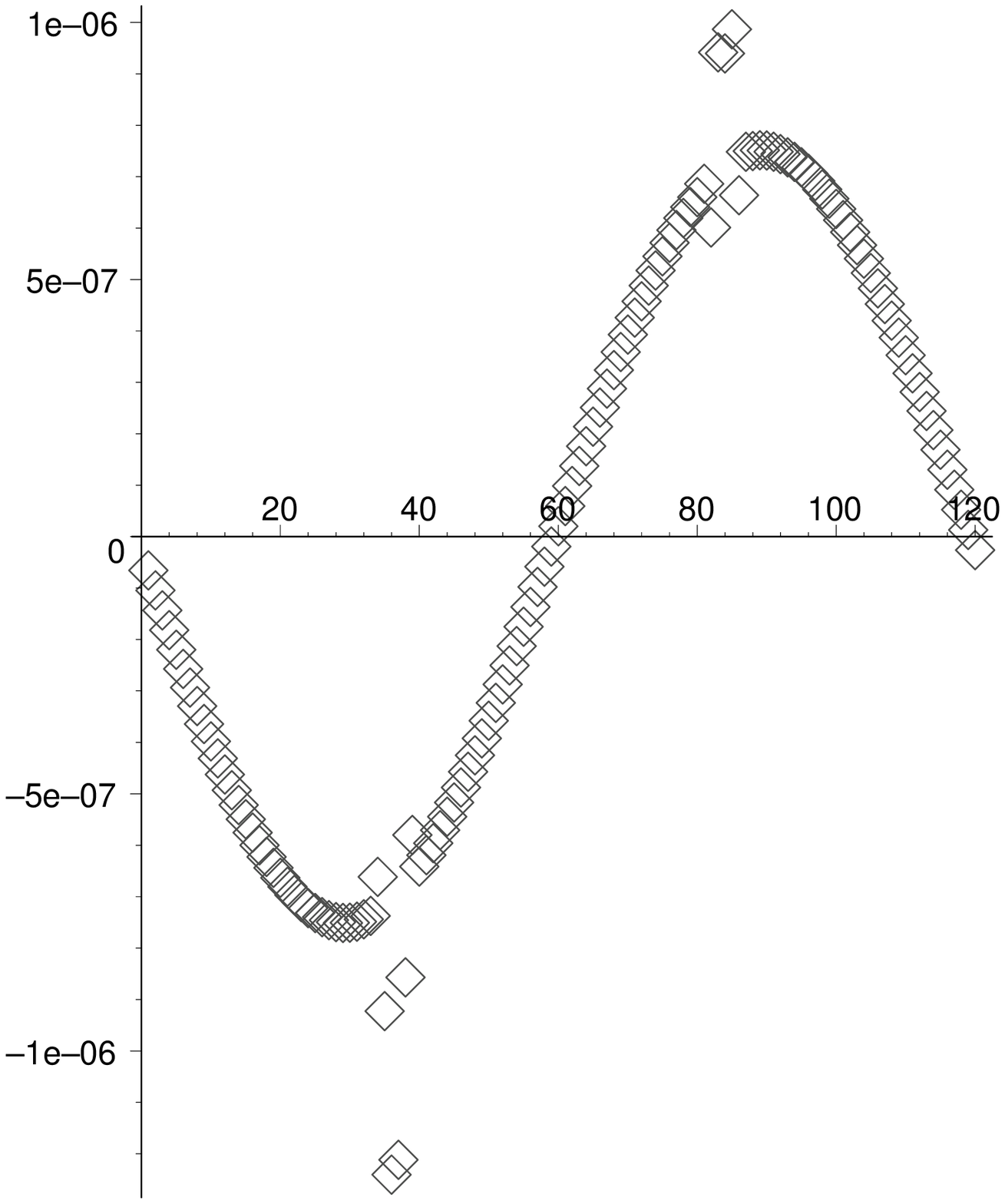}
\includegraphics*[height=7cm,width=5.5cm]{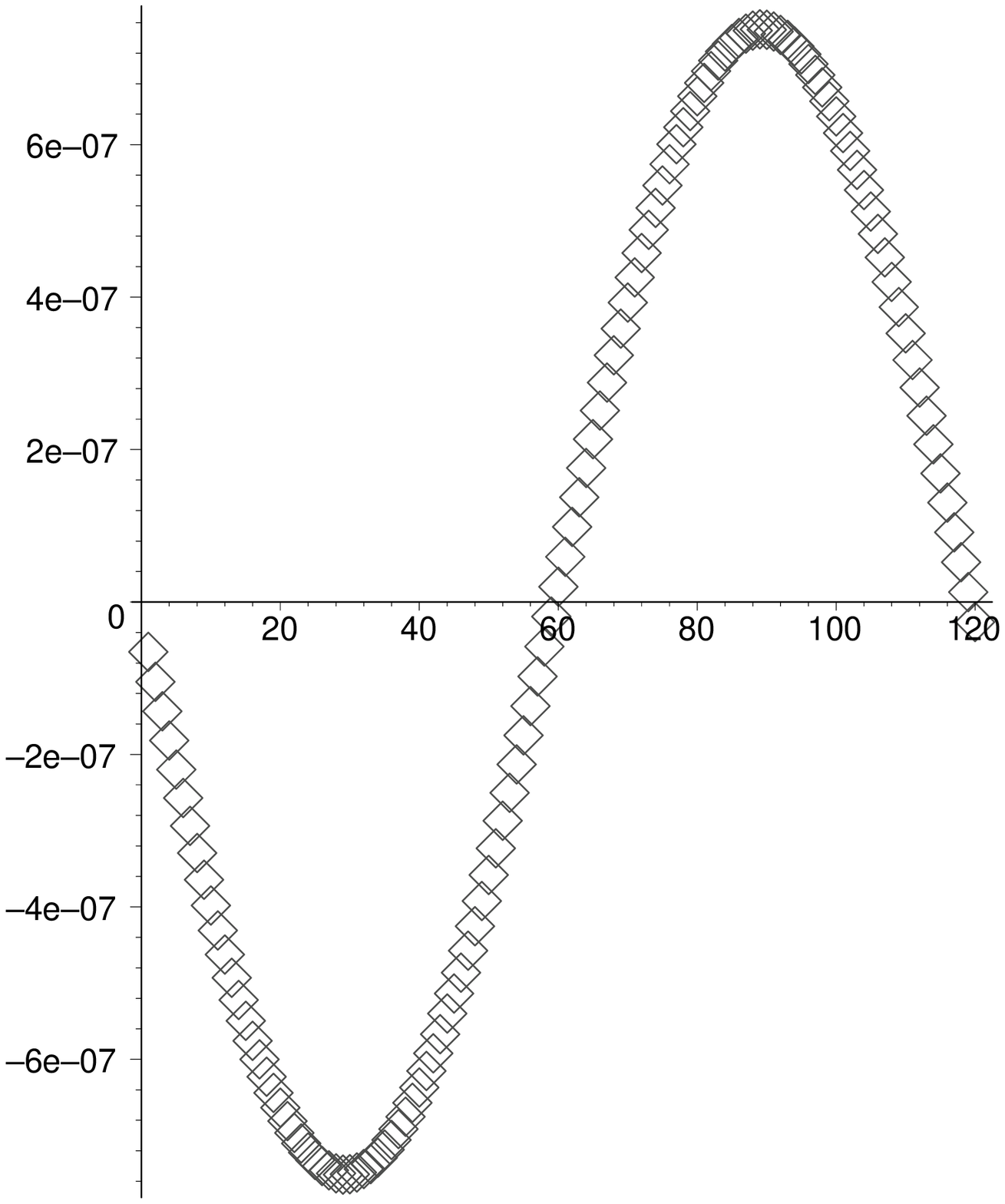}
\caption{Measured errors after one full step in the evolution of
$u=\sin(\pi x)$ in the domain $[0,1]$ under the three schemes considered 
---from left to right: the B-O, penalty and tapered schemes---. A CFL factor of
0.25 was employed and artificial dissipation of the Kreiss-Oliger form has
been added with $\sigma=0.1$. 
This value of $\sigma$ is not yet enough to smooth-out the errors in the B-O and
Penalty approaches, though in the first case makes the errors about the same
as those obtained with the tapered approach.}
\label{B4_error_diss0}
\end{center}
\end{figure}

\begin{figure}[ht]
\begin{center}
\includegraphics*[height=7cm,width=5.5cm]{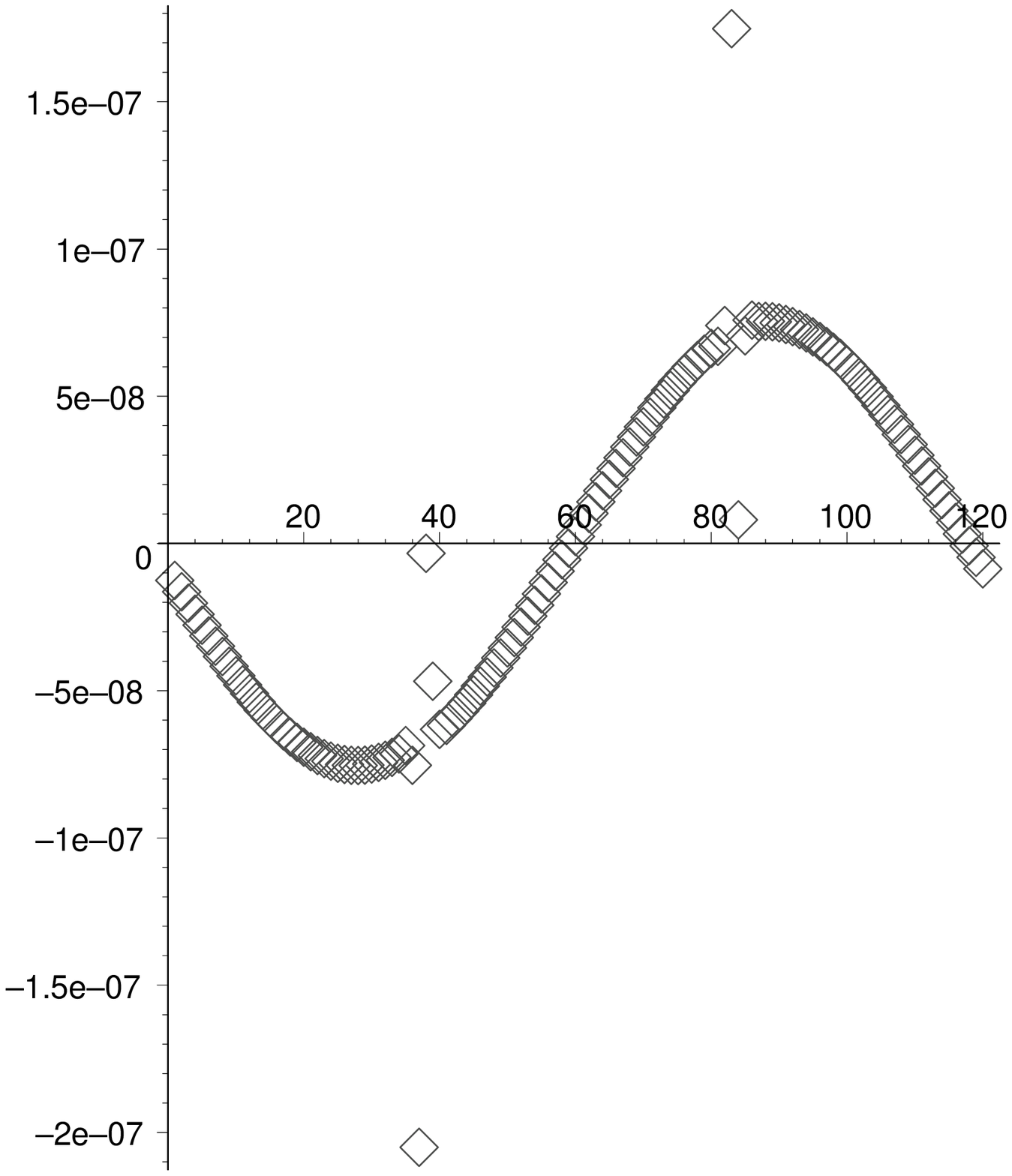}
\includegraphics*[height=7cm,width=5.5cm]{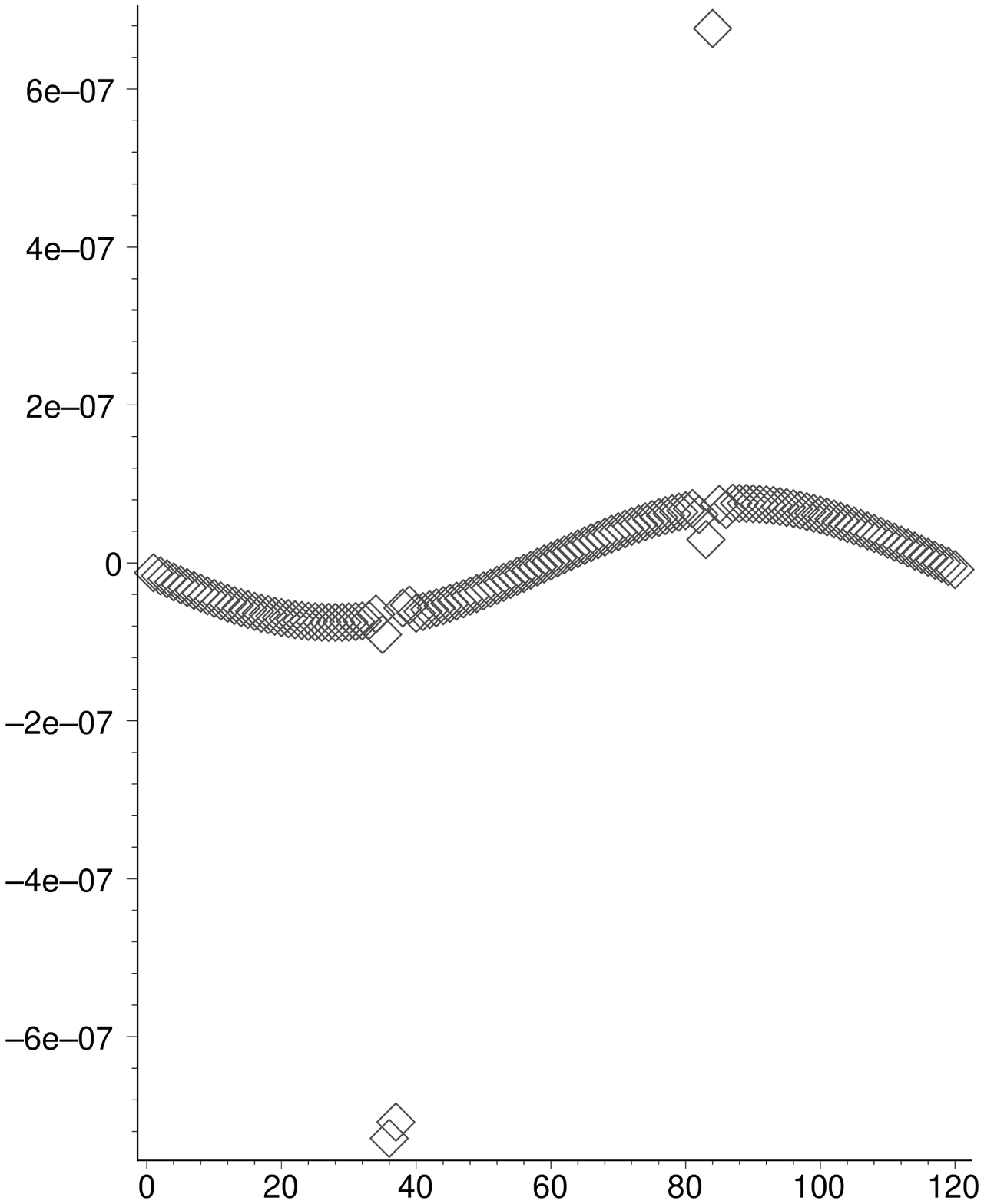}
\includegraphics*[height=7cm,width=5.5cm]{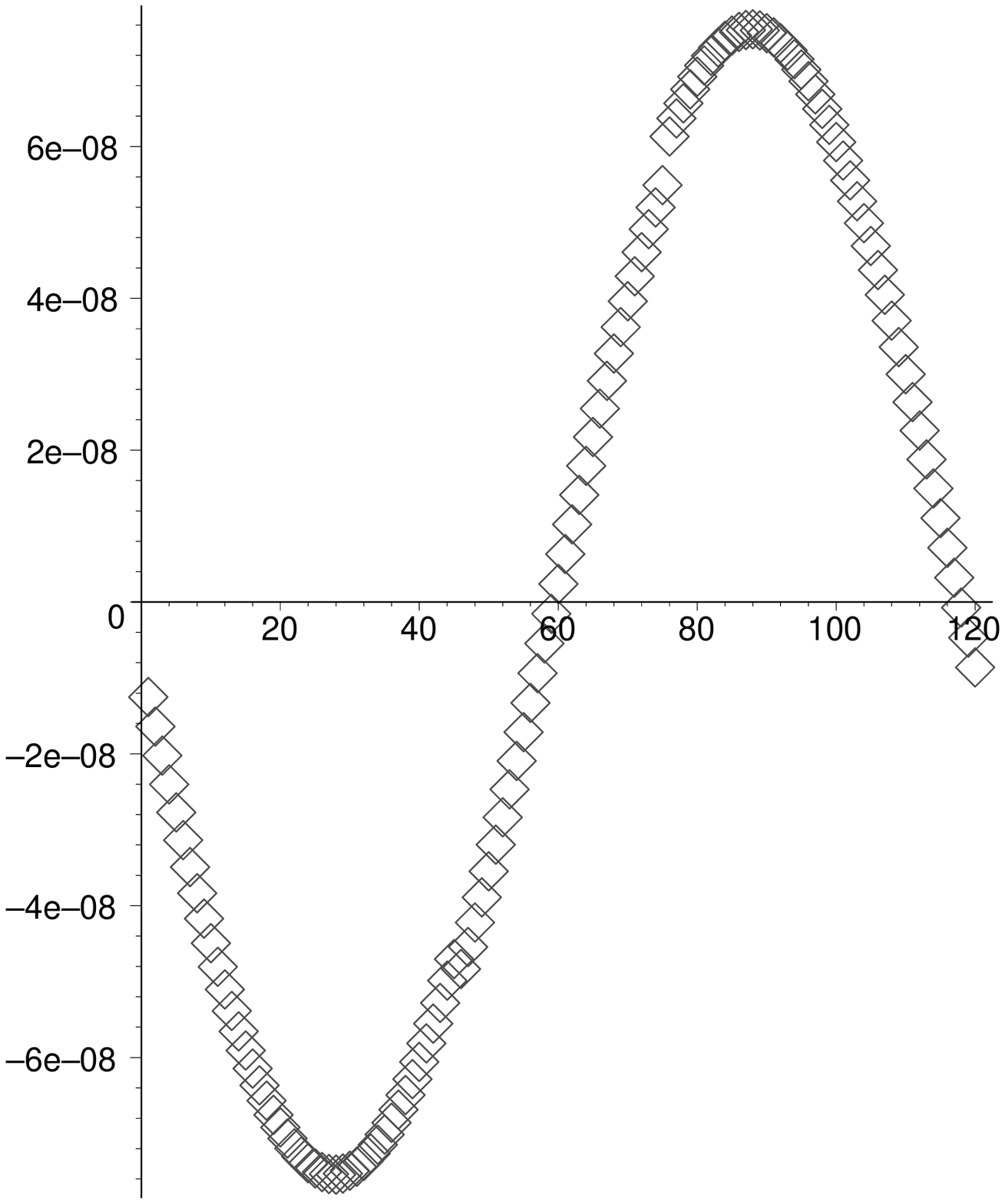}
\caption{Measured errors after one full step in the evolution of
$u=\sin(\pi x)$ in the domain $[0,1]$ under the three schemes considered 
---from left to right: the B-O, penalty and tapered schemes---. A CFL factor of
0.25 was employed and artificial dissipation Kreiss-Oliger type dissipation has
been added with $\sigma=0.01$. This value of $\sigma$ is too small to smooth-out the errors 
in the B-O and Penalty approaches while is still reasonable for the tapered approach.
Furthermore, the errors in the B-0 approach are now about $4$ times larger than in the
tapered one.}
\label{B4_error_diss1}
\end{center}
\end{figure}

As noted, the update scheme obtained for the BO and Penalty schemes are unstable for
the higher order case; this can be remedied or alleviated by the introduction
of some amount of dissipation. We consider the addition in two ways, one
as a dissipative operator modifying the right hand side of the equation (and hence
involved in all time levels), or as a high-pass filter only at parent levels.
The latter is amenable to be analyzed through the method outlined in Section~\ref{sec:analytical}
while the same for the former becomes too cumbersome. The analysis with the addition
of the filter shows that the addition of dissipation can render the scheme stable and experimental
tests (in the tests presented in the next section) indicates the same holds true with the first
option. Here we present results of the errors obtained after a single step in each of the
approaches considered, illustrated in figures \ref{B4_error_diss0} and \ref{B4_error_diss1}.
These figures show the errors obtained after a single evolution step of a (well resolved) smooth
initial data given by $u=\sin(\pi x)$. Figure \ref{B4_error_diss0} displays the observed behavior 
when no dissipation is employed. The errors obtained with the B-O and Penalty approaches are
about two orders of magnitude above those seen when employing the tapered approach after a single step.
The addition of dissipation alleviates this issue though the amount of the dissipation has to
be tuned to yield smooth errors while not affecting their magnitude. Figure \ref{B4_error_diss1}
show the obtained behavior with $\sigma = 0.1$ which is not yet enough to give smooth errors
for the B-O approach and even less for the Penalty method while quite suitable for the tapered approach.

\section{TESTS}
\label{sec:tests}
In the following one-dimensional series of tests, we consider four  natural
alternatives. These differ in their handling of refinement boundaries, in particular
how they compute derivatives there and what conditions are imposed.
Derivatives at interior points are defined by the standard centered ones.

These alternatives are: (I) sideways
derivatives (constructed so that their approximation order is equal to that of the 
centered derivative operator employed at the interior points); (II) derivatives 
of lower order near the boundaries, 
which, in the absence of a time interpolation, would satisfy summation by parts;
(III) Centered derivatives throughout, with extra ghost zones on each side of
the child grid as needed for applying the standard centered derivative. 
In the above three cases, boundary (and guard) data as the integration proceeds 
is obtained via interpolation from the parent grid.
The last option (IV) is the ``tapered boundary approach'' described in Section~\ref{sec:second}.

\subsection{Numerical Results. 1D tests}
We study the long term evolution of several systems implementing an FMR scheme.
Since in this case the grid-structure is fixed it does not 
correspond to a well
suited adaptive simulation, as the adaptivity would sequentially
create grids to follow the traveling features in the solution 
(assuming, of course, an appropriately defined refinement criterion is in place).
However, this arrangement is well
suited for our task here to assess the stability of the simulation and the
interaction of features in the solution propagating in and out of the finer
refined regions.

Our tests consist of:
\begin{itemize}
\item The first test is restricted to the simplest possible case provided
by the advection equation $u_{,t} = u_{,x}$. This is a hyperbolic system
with constant in time and space coefficients. For such a simple case,
it is observed that higher order accuracy can be achieved with a rather
simple interface boundary treatment.
\item The second test corresponds to a linearization of the Einstein equations
with respect to a time/space dependent background. 
\item The last test corresponds to the full Einstein equations restricted
to a spherically symmetric case. In this case, excision techniques are also
implemented.
\end{itemize}

\subsubsection{Advection equation}
We here concentrate on the advection equation $u_{,t} = u_{,x}$ discretized
in the domain $L_0 =\{x \in [-2,2] \}$ with two levels of refinement $\{L_1,L_2\}$ given by
$L_n =\{x \in [-1/n,1/n] \}$ ($n=1,2$). Periodic boundary conditions in the $L_0$ grid
are enforced directly at the difference operators level while interface boundaries $x=\pm 1/n$ are handled
via the standard B-O method or the tapered boundary approach. We implemented both
second and fourth order accurate derivative operators with a fourth order Runge Kutta
operator for the time integration.
The initial data is given by a compact support pulse given by
\begin{eqnarray}
u(t=0,x) = \left\{ \begin{array}{ll}
        \kappa (x/L-1)^5 (x/R-1)^5 & \mbox{if $x\in [L,R]$}  \\
        0 & \mbox{otherwise \, ,}
               \end{array}
        \right.
\end{eqnarray}
with $\kappa = 30$, $L=-0.55$ and $R=0.95$. These limits are arbitrarily defined
ensuring the pulse lies in the middle grid. As time progresses, the pulse will move
across all grids.

\begin{figure}[ht]
\begin{center}
\includegraphics*[height=8cm]{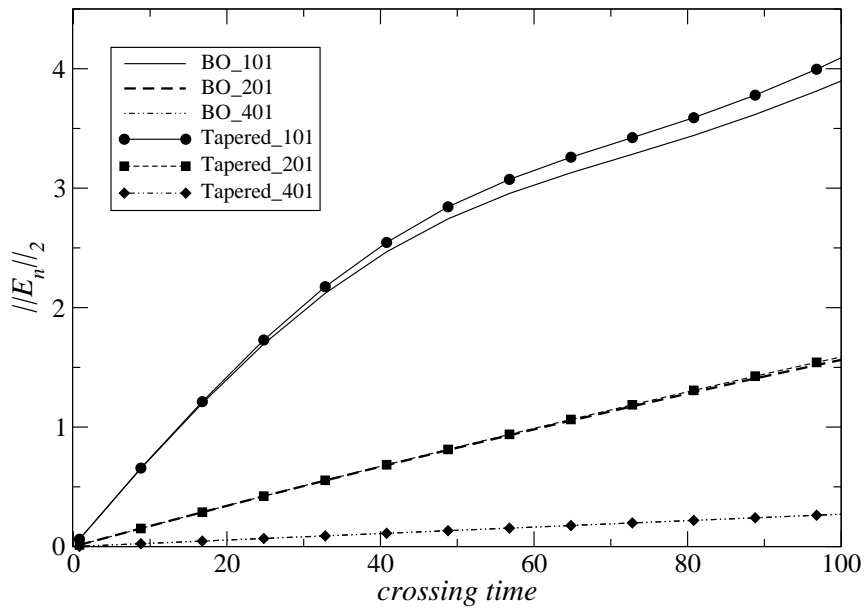}
\caption{Errors in the $L_2$ norm of the solution obtained with the B-O and tapered
boundary approaches. A CFL $=1 $ was employed and no dissipation. Note that the vertical
axis has not been rescaled the error in the solution indeed becomes very large after 
several crossing times manifested in large phase/amplitude errors.}.
\label{BOTAP_2nd_error}
\end{center}
\end{figure}

\begin{figure}[ht]
\begin{center}
\includegraphics*[height=8cm]{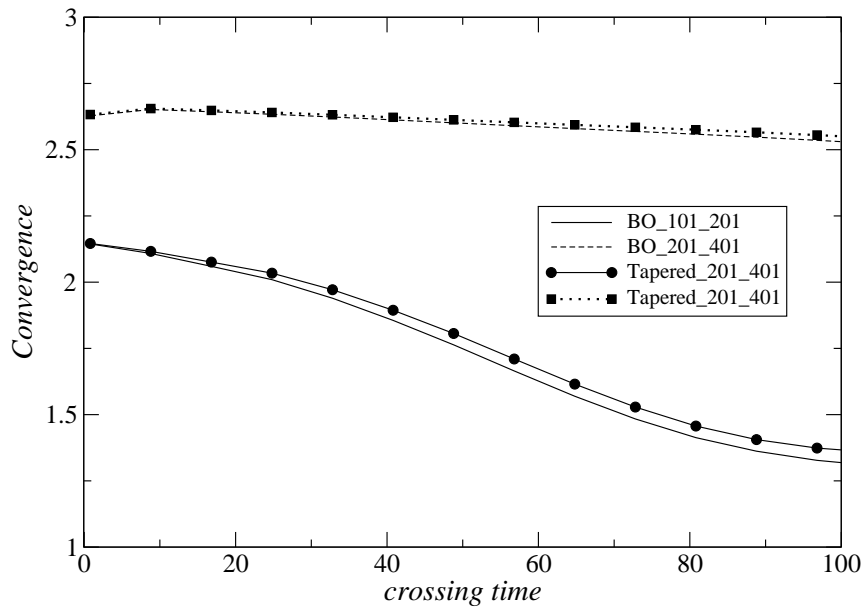}
\caption{Convergence rates measured for both cases obtaining similar results.}.
\label{BOTAP_2nd_conv}
\end{center}
\end{figure}

For the second order case, we have seen that all options discussed give a
stable implementation. We corroborated this by monitoring the error in the solution
obtained with both the BO and Tapered boundary approaches. To this end, we evolved
the solution on a uniform grid with $1601$ points and adopted it
as our ``analytical solution'' $F_a$ (recall that the unigrid problem is guaranteed to
be stable).
We then run with both implementations using a coarse grid consisting of $101,201$ and $401$ points
and calculate the error for each numerical solution $F_n$ as $E_n \equiv F_n - F_a$.
Figure \ref{BOTAP_2nd_error} illustrates the observed behavior which indicates both
approaches give practically the same solutions. Next we calculate the convergence rate
with these obtaining values consistent with second order accuracy as shown in figure~\ref{BOTAP_2nd_conv}.

When employing fourth order accurate derivative operators as discussed in
Section~\ref{sec:fourth}, boundary conditions are required in approaches (I) through (III).
We consider both linear and cubic interpolation employing in the first case
the values of the function at the coarser time levels and in the second case we
include the time derivative to obtain a cubical interpolant. These interpolations
provide boundary data at points  $x=\pm 1/n$ and
$x = \pm ( 1/n  -  1/(n (N_c -1)) )$. Notice that when employing
cubical interpolation in time and linear interpolation in space, no significant 
qualitative differences were observed if a higher order interpolation in space was employed.

As discussed previously, without the addition of dissipation, options (I) through (III)
yield unstable implementations, and we have confirmed this by monitoring
the $L_2$ norm of the solution vs. time. By running for several crossing times for different
resolutions, it is seen that the norm grows faster for finer refined grids.
Next, in order to control to some extent the instability, we add dissipation to 
the problem. This can be done in two forms, one is by modifying the right hand 
side of the equations by introducing a dissipative term and the other is to apply
a filter to the updated variables. 
In the second option the filter applied is, in a sense, equivalent to introducing
an integration on a fictitious time after each step (with respect to the coarsest
grid) of the equation $u_{,t} = \sigma d^3 \partial_{xxxx} u$. 
This results in the replacement
\begin{equation}
u^n_i \rightarrow u^n_i + \sigma \left ( 6 u^n_i -
 4 (u^n_{i+1} + u^n_{i-1}) + u^n_{i+2} + u^n_{i-2} \right ).
\end{equation}

\begin{figure}[ht]
\begin{center}
\includegraphics*[height=8cm]{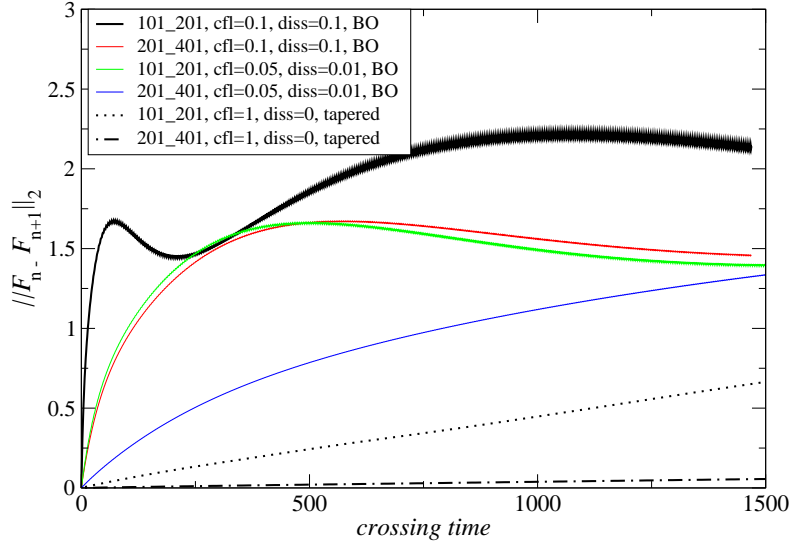}
\caption{Differences in the solutions obtained with different CFL/dissipation combinations}.
\label{BO_TAP_4th}
\end{center}
\end{figure}

\begin{figure}[ht]
\begin{center}
\includegraphics*[height=8cm]{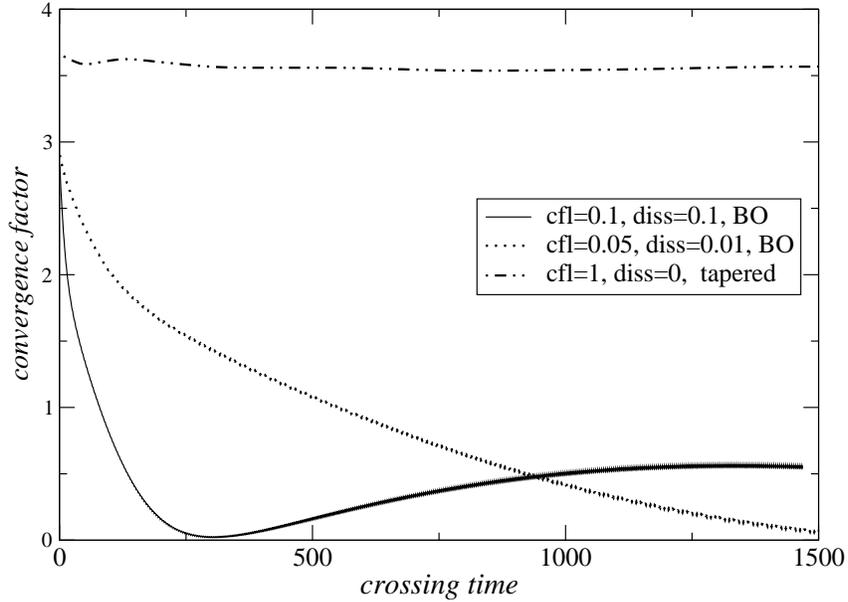}
\caption{Convergence factors obtained when employing different CFL/dissipation combinations}.
\label{BO_TAP_4thCONV}
\end{center}
\end{figure}

As a comparison of what is observed, figure \ref{BO_TAP_4th} illustrates the
the $L_2$ norm of the difference between the solution 
obtained with $N$ and $2N$ points vs. time (with $N=100,200$). The figure
illustrates the behavior of this quantity  for the solutions obtained with the B-O approach 
obtained with two different dissipation and CFL parameters. These solutions are obtained with 
strategy III (namely with two ghost zones on each side of the child grids to employ
the standard centered difference operator). We additionally show the corresponding
result for the tapered boundary approach.
As illustrated in the figure, while the dissipation manages to control the exponential
growth of the solution, convergence is severely affected.
The solution obtained with the tapered boundary approach on the other hand, behaves stably and
the error scales as expected.
Figure \ref{BO_TAP_4thCONV} shows the convergence factors calculated with these solutions,
while the tapered boundary approach yields a value quite close to the expected one of $4$,
the one obtained with strategy III quickly deteriorates.

\begin{figure}[ht]
\begin{center}
\includegraphics*[height=8cm]{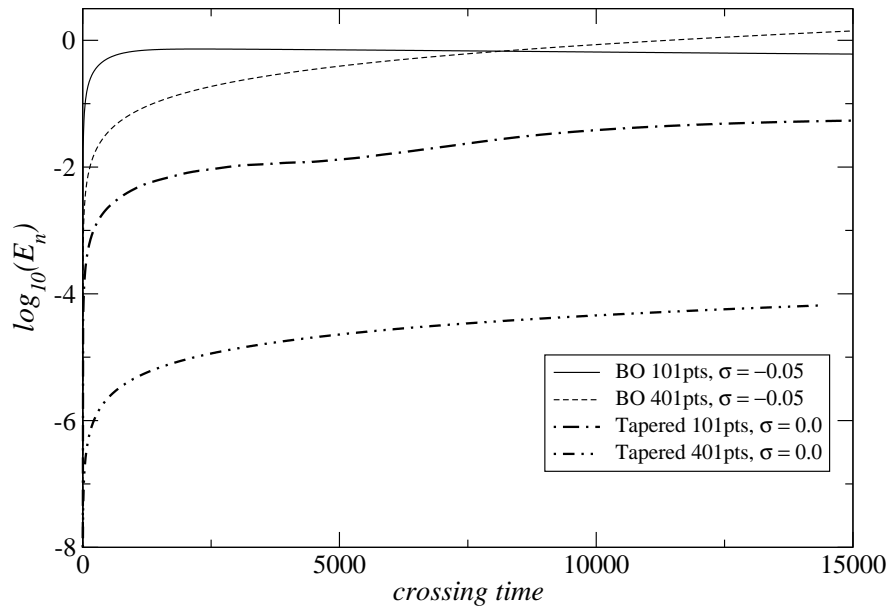}
\caption{Errors in the $L_2$ norm of the solution obtained with the B-O and tapered
boundary approaches. A CFL $=1$ was employed and $\sigma = -0.05$ in the B-O case.}.
\label{BO_TAP_4thERROR}
\end{center}
\end{figure}

Finally, as was discussed in Section~\ref{sec:fourth}, unless both the CFL and 
dissipation parameter are chosen
carefully, this filtering might not be enough to control the instability.
For instance, for a CFL factor $=1$ and $\sigma = -0.05$, the growth of the instability
is less severe and might not be visible for relatively coarse
resolutions for long times though as the grid is refined 
the instability might become evident.
We illustrate this behavior, and contrast it with the one observed with the
tapered grid approach, by monitoring the $L_2$ norm of the error $E_n$ defined as in
the second order case (i.e. with respect to the solution obtained with a unigrid evolution
on a grid consisting of $3201$ points). Figure \ref{BO_TAP_4thERROR} shows the results obtained
for the coarse grids having $101$ and $401$ points. Clearly, while the tapered boundary approach
gives the expected behavior, strategy (III) with the addition of dissipation fails to give
good answers. Notice that while the solution obtained with $101$ points does not exhibit 
an instability the one with $401$ points does so clearly. The reason behind this is that
the instability for the coarser case has not yet grown above the truncation error in the solution.

\subsubsection{Linearized Einstein equations}
The second test is defined by considering the linearized
Einstein equations off the ``gauge-wave'' solution. This solution describes
flat-spacetime in a coordinate system that introduces an explicit time-position
dependence of the metric as
\begin{equation}
ds^2 = e^{A \sin(\pi (x-t))} (-dt^2 + dx^2) + dy^2 +dz^2
\end{equation}
We then write the linearized Einstein equations off this spacetime  in a first
order symmetric hyperbolic form as discussed in \cite{multipatch1} following the formulation introduced
in \cite{sarbachtiglio} which includes a dynamical lapse. By restricting to
perturbations in the $t,x$ directions, solely depending on these coordinates, one obtains a reduced system
of non-trivial equations for $\delta g_{xx}, \delta K_{xx}, \delta d_{xxx}, \delta \alpha, \delta A_x$.
We solve the resulting equations in the domain $x\in [-1,1]$, with $A=0.1$ and compare
the results obtained with 2nd and 4th order accurate spatial derivative operators while maintaining
$4th$ order accurate integration in time via a Runge Kutta algorithm.

We first run a series of tests employing a $2nd$ order accurate spatial derivative operators, a 4th order interpolation
to create child grids, a $4th$ order Runge-Kutta
time integrator and a direct injection from child to parent grid points. We concentrate on
the accuracy and convergence rate observed for cases (I-III) and (IV) above. Ie, in the first 
case we employ second order derivative
operators at interior points while first order one at the boundaries. Since boundary points
in this case are over-written by the time-interpolation cases (I-III) are identical.
For case (IV) however, we define extra child points via interpolation from parent to child
at level $n-th$. These extra points are discarded at the $n+1th$ level and are solely used 
to remove any dependence on the refinement boundaries that child points, at the region of interest, 
have at the $n-th$ level.

\begin{figure}[ht]
\begin{center}
\includegraphics*[height=8cm]{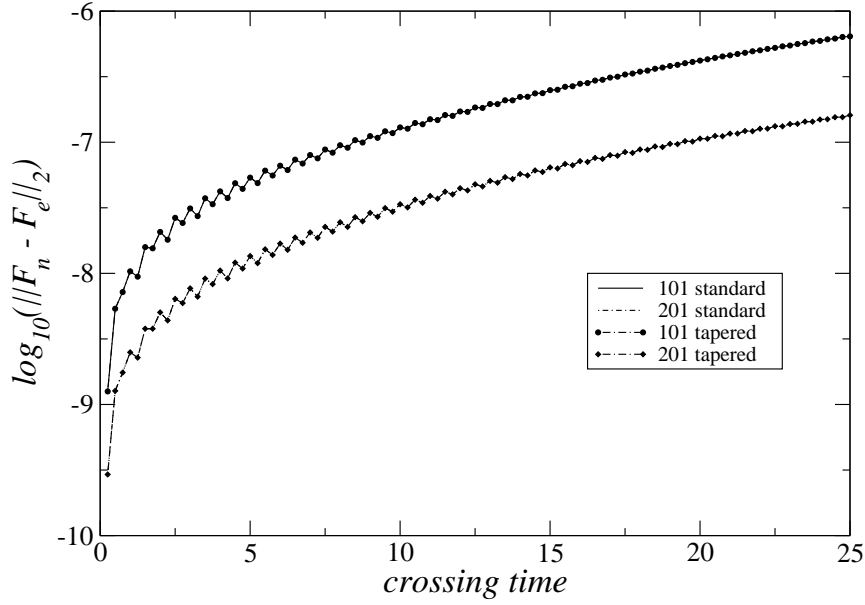}
\caption{Errors associated with cases (I-III) and (IV) as a function of time. The lines are 
indistinguishable from each other.}
\label{2ndordercompareGW}
\end{center}
\end{figure}

\begin{figure}[ht]
\begin{center}
\includegraphics*[height=8cm]{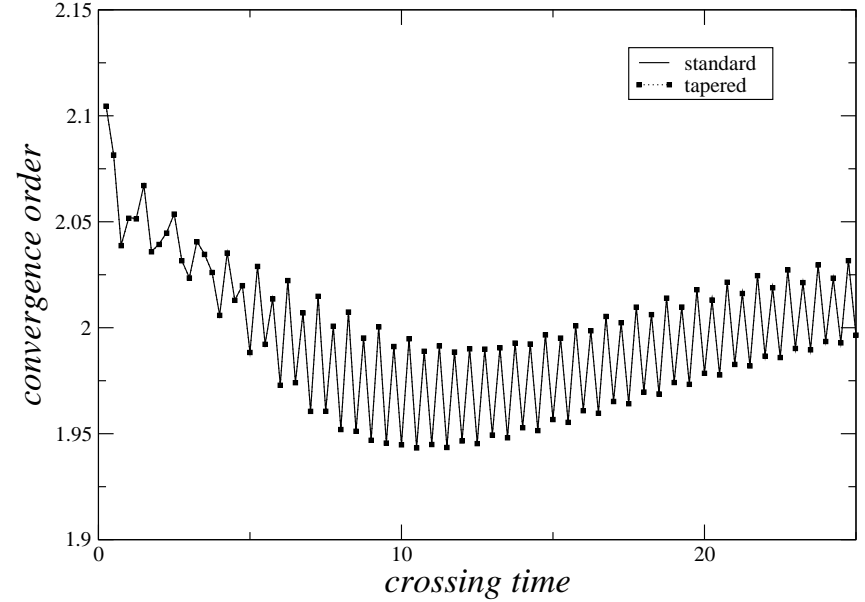}
\caption{Measured convergence rate for the different methods. There is no appreciable
difference in the 2nd order case. Again, there is no appreciable difference with
the values obtained with both approaches.}.
\label{2ndorderconvergeGW}
\end{center}
\end{figure}

First we confirmed that the unigrid code is
stable and convergent, and pick the solution obtained with the highest resolution
(corresponding to a grid with 3201 points) and consider it the `analytically' expected 
solution $F_a$ which we use to compare the results obtained when using fixed mesh refinement. 
For these tests we employ a base grid of $101$ and $201$ points, each with 2 levels of
refinement (with boundaries at $[-1/q,1/q]$ with $q=2,4$) and no dissipation
is included. Figure \ref{2ndordercompareGW} illustrates the (log of the) $L_2$ norm of 
the difference between the solution obtained at the base grid $F_n$ and
 $F_a$ with $n=101,201$. The obtained results with the standard
FMR approach and the tapered boundary one are practically indistinguishable from each other.
This also implies the convergence of both should be practically the same, ie. second order,
 a fact that is shown in figure \ref{2ndorderconvergeGW}.

In the $4th$ order accurate case, as expected from the analysis presented
in section II, differences do arise. For these tests we employ 4th order
derivative operators at interior points --suitably modified near the
boundaries to cover cases (I) and (II) or ghost zone points suitable 
added for case (III)--; a 6th order interpolation from parent
points to define child grid points; a 4th order RK time integrator and a direct
injection from child to parent grid points.
As before, we first run a series of 
unigrid tests to confirm that a stable evolution is obtained as expected and
choose the solution obtained with the highest resolution run (obtained with a grid consisting of 
3201 points) and consider it the `analytically' expected solution $F_a$.
Again we consider a base grid of $101$ and $201$ points, each with 2 levels of
refinement (with boundaries at $[-1/q,1/q]$ with $q=2,4$) and as above no dissipation
is included. 

\begin{figure}[ht]
\begin{center}
\includegraphics*[height=8cm]{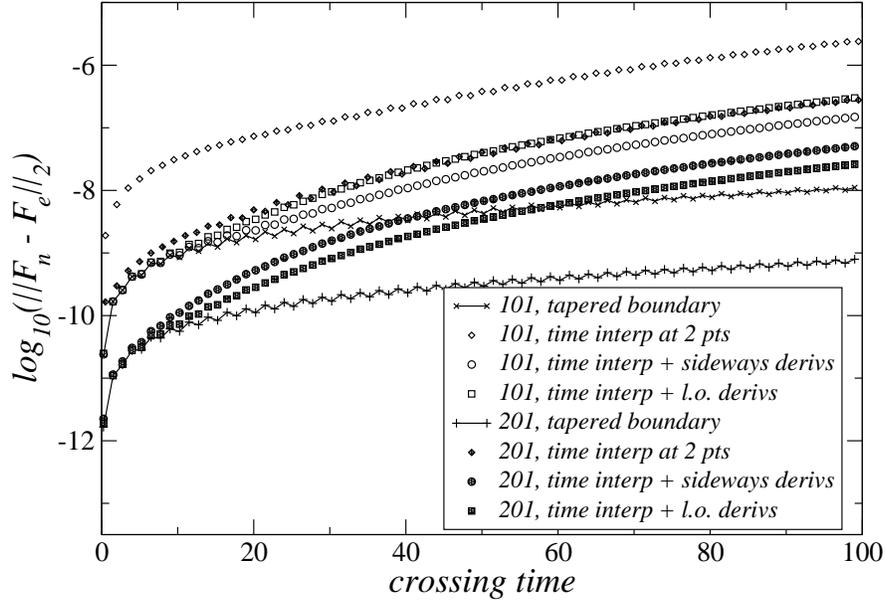}
\caption{Errors associated with the 4th order accurate approaches. The errors
associated to the cases where time interpolation is employed grows exponentially
fast while those measured in the tapered boundary approach do so linearly.}.
\label{4thordercompareGW}
\end{center}
\end{figure}

\begin{figure}[ht]
\begin{center}
\includegraphics*[height=8cm]{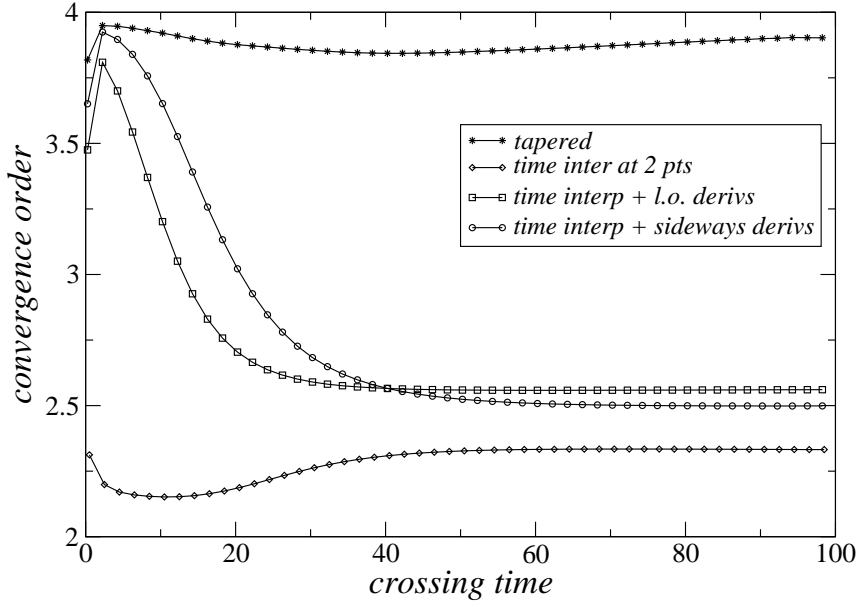}
\caption{Measured convergence rate for the different cases, while the tapered
boundary approach remains close to 4 over 100 crossing times, the convergence
of the other methods fall sharply.}.
\label{4thorderconvergeGW}
\end{center}
\end{figure}

Figure \ref{4thordercompareGW}
illustrates the (log of the) $L_2$ norm of the difference between the ``analytical'' solution
and those obtained with a coarse grid (plus the two refinement level) $F_n$ with $n=101,201$. 
Clearly the errors associated
with the implementations (I-III) grow considerably faster (indeed exponentially) than those
obtained with the tapered boundary approach (linearly). In fact, after about $50$ crossing times
the error associated with the tapered boundary approach using $101$ base points is approximately
the same as that with the standard approach but with a base grid of $201$ points. 
Figure \ref{4thorderconvergeGW} illustrate the measured order of convergence for
these methods, while the value obtained with the ``standard approaches'' significantly 
reduces as time proceeds the corresponding one to the tapered boundary approach remains
near $4$ decreasing just slightly as errors accumulate over time.

\begin{figure}[ht]
\begin{center}
\includegraphics*[height=8cm]{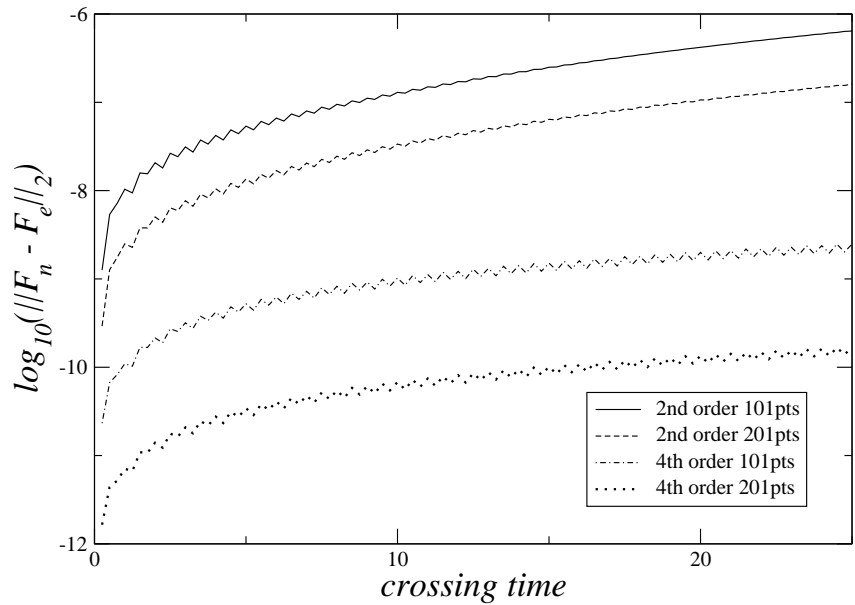}
\caption{Errors associated with the 2nd and 4th order tapered boundary approaches.}.
\label{4th2ndnew}
\end{center}
\end{figure}

Last, for illustrative purposes, we show in figure \ref{4th2ndnew}
the errors associated with the different solutions obtained with
the tapered boundary approach for $2nd$ and $4th$ order accurate spatial operators.
The accuracy gained by employing a higher order operator is evident.

\subsubsection{Spherically symmetric spacetime}
In the last one-dimensional test we consider Einstein equations restricted
to spherically symmetric spacetimes in vacuum. The solution for a black hole
initial data must be static and hence field variables have 
only non-trivial radial dependence. 

We again employ a grid structure composed of a base grid given by
$L_0 =\{r \in [M,11M] \}$ with two levels of refinement $\{L_1,L_2\}$ given by
$L_n =\{r \in [M,11/(2n) M] \}$ ($n=1,2$). Thus, while the boundaries at the
right boundaries of each domain are sequentially located at smaller radii, the
left boundaries all coincide at $r=M$. At these boundaries, for the tapered-boundary
approach one takes advantage of the fact that all characteristic variables
are inflow, and hence the values at each grid at and near these boundaries
can be updated using solely grid values within that grid.

\begin{figure}[ht]
\begin{center}
\includegraphics*[height=8.5cm]{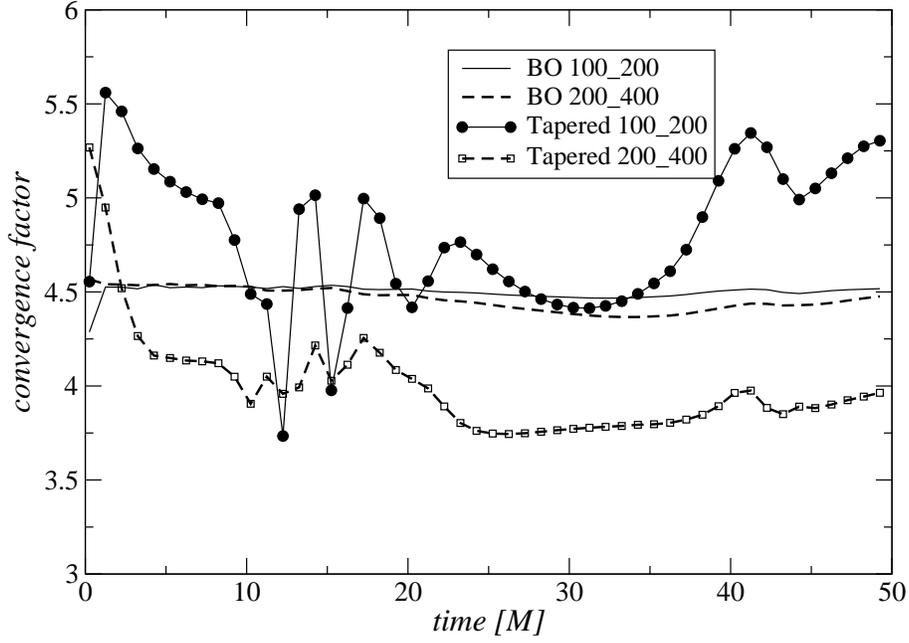}
\caption{Convergence factors obtained with the numerical solution of $g_{rr}$ for both
the BO and tapered boundary approaches. In both cases, a behavior consistent with fourth order
accuracy is obtained.}.
\label{conv_grr}
\end{center}
\end{figure}

We implemented Einstein equations in 1+1 dimensions as described in \cite{KST,cpbc} with
fourth order accurate derivative operators  at interior points (suitably modified near
boundary points as dictated by the SBP requirement) and a fourth order Runge-Kutta update scheme in time.
We adopt unperturbed initial data corresponding to Schwarzschild in ingoing
Eddington Finkelstein coordinates with an analytically given shift and the
free part of the densitized lapse function. At the outer boundary of the coarse grid,
constraint preserving boundary conditions are imposed as described in \cite{cpbc}
while the artificial boundaries are either handled via the standard B-O scheme (interpolating
two ghost-zone points) or via the tapered grid approach. A small amount of dissipation
is added in all tests with a fifth order Kreiss-Oliger style operator $\sigma h^5 (D_- D_+)^3$
suitably modified near the boundaries as explained in \cite{multipatch1} (in our runs $\sigma=0.01$).
Since the solution should converge to the analytical value, we can compute the error with
respect to it. Figure \ref{conv_grr} illustrates  the convergence factor obtained from the
variable $g_{rr}$ (for $100, 200$ and $400$ points) in the coarse grid. For both cases the 
obtained convergence factor is consistent with the expected value of $4$. 

\begin{figure}[ht]
\begin{center}
\includegraphics*[height=8.5cm]{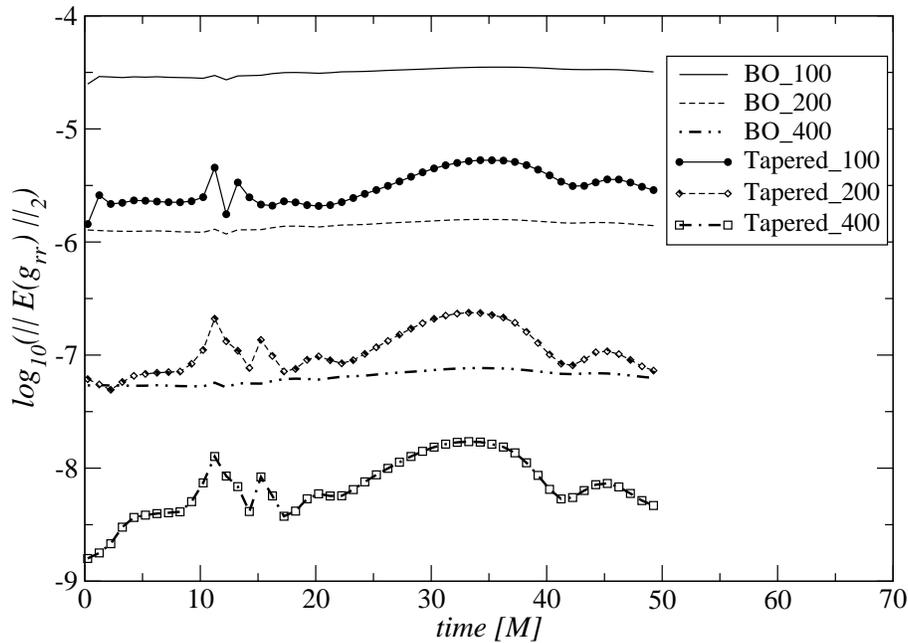}
\caption{Convergence factors obtained with the numerical solution of $g_{rr}$ for both
the BO and tapered boundary approaches. In both cases, a behavior consistent with fourth order
accuracy is obtained.}.
\label{errors_grr}
\end{center}
\end{figure}

Last, we examine the behavior of the errors themselves for the resolutions employed. Figure \ref{errors_grr}
display the $L_2$ norm of the errors $E(g_{rr})\equiv g_{rr}^n - g_{rr}^a$ with  $g_{rr}^a,g_{rr}^n$ the analytical
and numerical solution (the latter obtained with $n$ points in the coarse grid). Notice that although
both schemes yield a stable evolution, the errors associated with the tapered grid approach are 
about an order of magnitude smaller than those obtained with the BO scheme.

\subsection{Numerical Results. 3D tests}
\subsubsection*{Semi-linear wave equation}

To test these ideas in three dimensions, we use an implementation of the
semilinear wave equation. A scalar field $\phi$ obeys the wave equation
\begin{equation}
\Box \phi = Q \phi^P
\label{sll:eom}
\end{equation}
for odd $P$ and where $Q \in \left\{ -1, 0 , 1 \right\} $ serves to parameterize the
nonlinearity. For $Q=0$ the linear case is obtained, whereas for $Q=+ 1$ the
model is focusing (studied in~\cite{lieblingAMR})
and for $Q=-1$ the model is defocusing. Choosing Cartesian 
coordinates and defining as the fundamental fields in order to cast the problem
in first order differential form:
\begin{eqnarray}
\Pi(x,y,z,t)    & \equiv & \frac{\partial}{\partial t} \phi \\
\Phi_x(x,y,z,t) & \equiv & \frac{\partial}{\partial x} \phi \\
\Phi_y(x,y,z,t) & \equiv & \frac{\partial}{\partial y} \phi \\
\Phi_z(x,y,z,t) & \equiv & \frac{\partial}{\partial z} \phi,
\end{eqnarray}
the equation of motion Eq.~(\ref{sll:eom}) takes the form of the following system of
equations
\begin{eqnarray}
\dot \phi   & = & \Pi\\
\dot \Phi_x & = & \Pi_{,x} \\
\dot \Phi_y & = & \Pi_{,y} \\
\dot \Phi_z & = & \Pi_{,z} \\
\dot \Pi    & = & \left( \Phi_x \right)_{,x}
                 +\left( \Phi_y \right)_{,y}
                 +\left( \Phi_z \right)_{,z}
                 + Q \phi^P.
\end{eqnarray}
Solutions of this system are found by incorporating appropriate finite differences
within a distributed AMR infrastructure. Using fourth order accurate finite difference
approximations of spatial derivatives along with a Runge-Kutta time integrator accurate
to third order, convergence to third order is expected as the resolution is
increased. This serves also to illustrate that the tapered grid approach allows for
decoupling the order of the time-integrator and the derivative operators. 

\begin{figure}[ht]
\begin{center}
\includegraphics*[width=13cm]{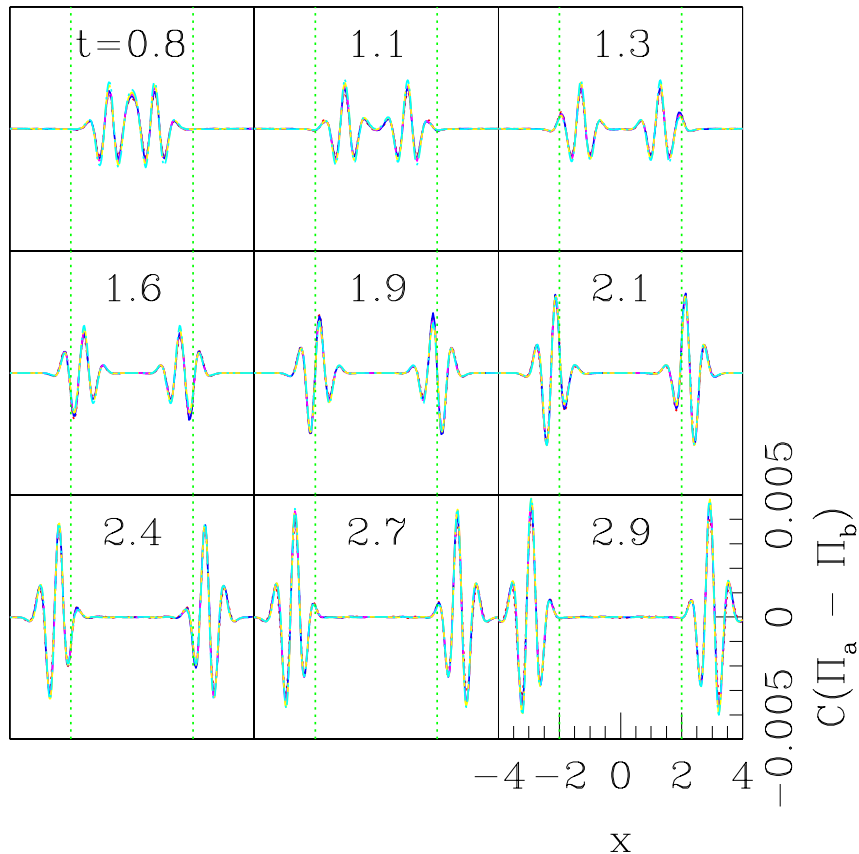}
\caption{
   Shown are the differences between runs of various resolutions rescaled
   according to what would be expected for third order convergence,
   $C \left( \Pi_a - \Pi_b \right) $. The initial data consists of
   an initially static Gaussian pulse centered about the origin for the case where $Q=0$.
   That
   the rescaled differences agree, even as the pulse traverses the
   boundary of the refined region (the green vertical lines), suggests
   that third order convergence is maintained:
   red,     dotted           line ( $\Pi_{145}$ - $\Pi_{181}$), $C=1$,
   blue,    solid            line ( $\Pi_{181}$ - $\Pi_{217}$), $C=2.26$, 
   magenta, short dashed     line ( $\Pi_{217}$ - $\Pi_{253}$), $C=4.45$, 
   yellow,  long dashed      line ( $\Pi_{253}$ - $\Pi_{289}$), $C=7.92$, and
   cyan,    dot-short dashed line ( $\Pi_{289}$ - $\Pi_{325}$), $C=13.1$.
}
\label{sll:conv}
\end{center}
\end{figure}

An example of such a convergence test is shown in Fig.~\ref{sll:conv}. A coarse
grid with bounds in each direction of $\pm 4$ is established with a (fixed mesh)
refined  ({\tt 2:1} refinement) region established with bounds of $\pm 2$.
Evolutions were carried out for increasing resolution and the results compared.
The lines shown indicate the rescaled differences
$C \left( \Pi_a - \Pi_b \right) $ where $a$ and $b$
indicate the number of points in
each dimension, $x$, $y$, and $z$.
The rescaling factor $C$ is computed as
$C \equiv [ 1/(145-1)^3 - 1/(181-1)^3 ] / [  1/(a-1)^3 - 1/(b-1)^3 ]$ which
is how one expects these differences to scale if indeed the code is third order
convergent. As the results in the figure show, these differences do decrease
as expected. Furthermore, although the size of the differences increase in absolute
terms as the pulse propagates across the refinement boundary, convergence and
smoothness is not disturbed.

\begin{figure}[ht]
\begin{center}
\includegraphics*[width=13cm]{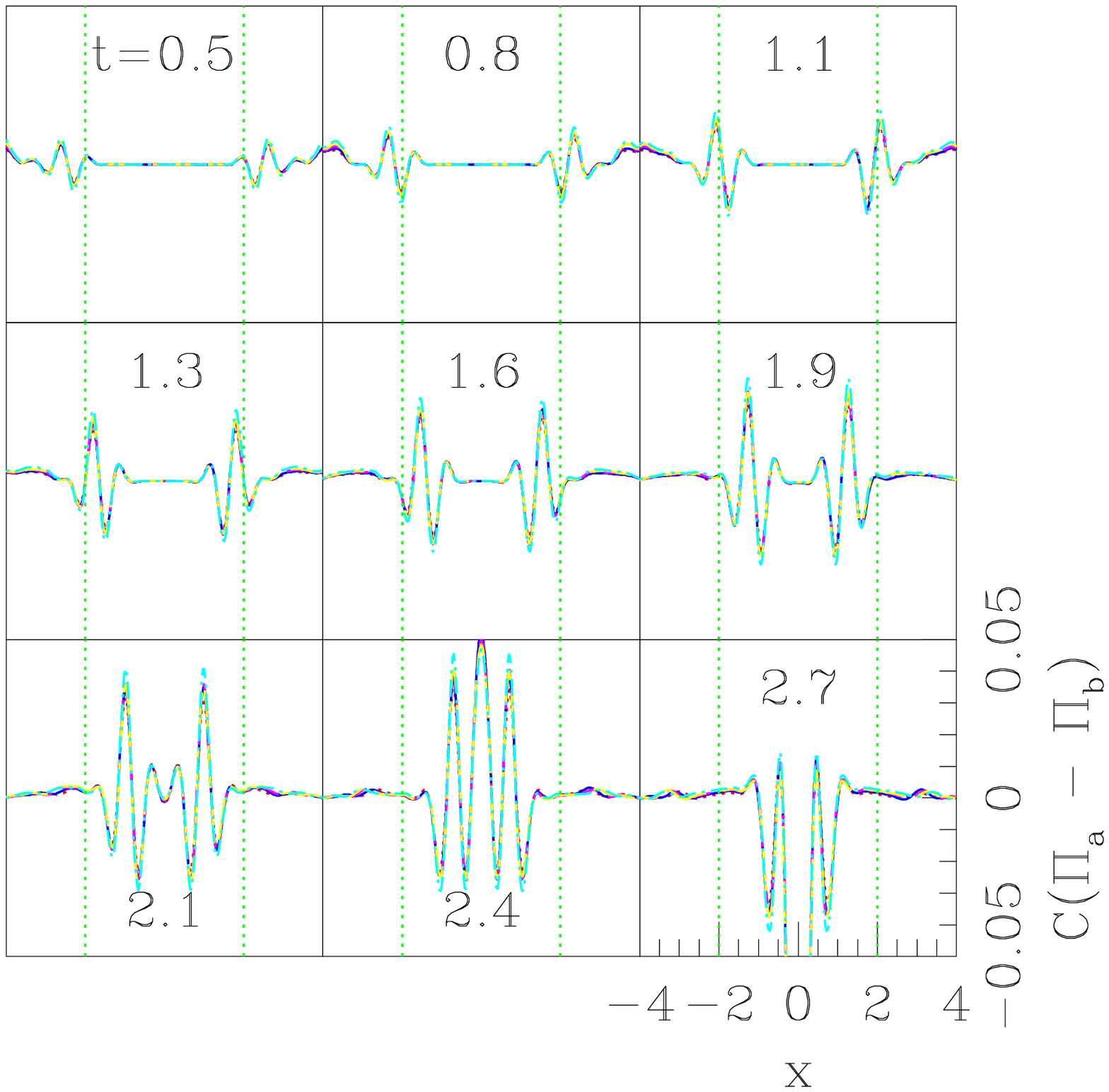}
\caption{
   Shown are the various rescaled differences between runs of various resolutions
   for a nonlinear case ($Q=+1$, $P=7$) of an ingoing wave pulse.  The resolutions are the same
   as those shown in Fig.~\ref{sll:conv}, and similar agreement among them suggests 
   third order convergence. The evolutions demonstrated singular behavior soon after the
   last frame, placing this evolution confidently within the nonlinear regime of the model.
}
\label{sll:conv2}
\end{center}
\end{figure}

Similar results are obtained for a nonlinear example, as shown in Fig.~\ref{sll:conv2},
where a nonlinear case ($Q=1,P=7$) is employed. The initial data describes
and incoming pulse which display similar behavior at $t \simeq 2.75$. Our simulations 
show that even in this highly non-linear regime, the solutions converge to the expected
order.

\section{Conclusion and final comments}
\label{sec:conclusion}
In the present work we have presented an analysis of different alternatives
aiming to yield a stable FMR/AMR implementation. It is observed that
the straightforward application of the standard B-O scheme has unstable
modes and that natural variations of this scheme share this problem.
On the other hand, the tapered boundary approach, in which points that have
been affected by the interface boundary treatment are discarded, is clearly stable.
In addition to being free of unstable modes, by discarding these points the
reflections at the interface between child and parent grids are significantly
reduced. 

The unstable modes found are represented by eigenvalues of the amplification
matrix with norm larger than $1$. An inspection of these reveal that
they differ from unity by a small amount (for the toy model considered, and typical
grid/CFL values are $\simeq 1.0005$). Therefore, the instabilities associated with 
these modes will grow slowly for few refinement levels and might not
become too evident for evolutions of moderate length. It is thus not surprising that in 
common applications in numerical relativity where typical CFL factors employed are of order unity,
relatively coarse resolutions are employed and the update scheme used is
the so-called iterative Crank Nicholson ---which is inherently much more dissipative than the
Runge-Kutta method considered here--- simple test-beds with few levels of refinement do not
display unstable modes.  For several
refinement levels, the instabilities will play a role unless suitably dissipated away.
However, we stress again that even if the instabilities might be controlled this way, the
errors obtained in all our tests with ``standard'' approaches are significantly larger than those
seen when employing the tapered grid approach.

Other schemes have been recently introduced to reduce the interface reflections for
second order methods\cite{bakerAMR}.
This involves suitably modifying the derivative operators to preserve smoothness in the error of
the advection equation at the interfaces. Although the application of these operators 
give improved results in black hole spacetimes, the results presented did not include sub-cycling
and higher derivative schemes have not yet been worked out.

Finally, although we have noticed that the penalty approach does not help in developing a stable
AMR strategy, it does play a key role in the stable handling of multiple grid structures
that abut~\cite{multipatch1}.
Thus an obvious strategy to implementing AMR techniques within a multiple grid structure is to combine
both the penalty technique with the tapered approach. The penalty technique would be used to deal 
with different grid interfaces
while the tapered approach to handle artificial interface boundaries as presented in this work.

Throughout our work we have restricted ourselves to fully first order equations because,  in this case,
a number of results exist on how to ensure stability of the fully discrete implementation.
Such results are not available for fully second order or first order in time/second in space systems,
though work in this direction is underway\cite{kreiss2ndorder,calhinderhusa}. Certainly, the tapered
approach can still be used in both these types of systems as it only relies on having a stable unigrid implementation
and ensuring enough fine grid points in the numerical past domain of dependence at level $n$ to
define the solution at level $n+1$.

In summary, the tapered grid approach provides a safe and natural way to build a stable AMR/FMR scheme
of arbitrary order. It has a non-trivial overhead cost, but this is offset by the significantly
smaller errors and spurious reflections the solutions obtained with it has when compared to
existing alternatives.
Note: while completing this work a scheme, similar in spirit, for a fourth order implementation has 
been presented in \cite{Csizmadia:2005zq} though in this case time and space interpolations
are used to fill needed points while the boundary is tapered off.

\appendix
\section{Comparison cost of two `stable' FMR alternatives: 2nd order B-O vs. 4th order tapered}
\label{sec:cost}
We here compare two stable options for the $u_{,t} = u_{,x}$ problem
in their associated computational cost. We concentrate on 
the relatively inexpensive BO 2nd order scheme and
a tapered boundary approach implemented with 4th order accuracy.

The particular issue we concentrate on is whether the additional
expense introduced by the 4th order tapered boundary approach renders
it too costly to reach a target error when compared with a 2nd
order implementation.

To obtain a conservative estimate we assume a base grid with $M+1$ points covering
a cubical domain of size $L$ and so we define $h\equiv L/M$.
We assume that within this domain, a further refined grid is needed which
has $N \simeq M$ points.

Now, suppose one has the option of adopting a 2nd order or a 4th order
accurate overall stable scheme. We want to estimate the cost that each one
will have associated to it. The computational cost will be determined by
the cost of the coarse and fine grid updates. For the coarse grid one
needs to take second or fourth order derivatives (which involves two and six
operations per point respectively) and the update step (which we take as Runge-Kutta
3rd and 4th order respectively). The cost to update the coarse grids is simply
governed by the derivative calculation and the Runge-Kutta update. A simple
calculation indicates the cost is $23$ ($38$) floating point operations 
per point for the second (fourth) order
derivative operators plus Runge Kutta 3rd order (4th order). These result from adding the
cost associated to computing the different Runge-Kutta steps (ignoring the storage cost),
to that of the final update.

The algorithm to update the child grid has four basic steps: {\bf A} creation
of the child grid, {\bf B} computation of derivatives, {\bf C} update and {\bf D} injection
from child to parent grid. The cost associated with the child grids for {\bf B} and
{\bf C}, --per point--, are the same as those in the parent grid. However, since for
the 4th order case the child grid must be enlarged, we will here consider
that the number of child grid points doubles (hence there will be $N$ and $2N$ points
associated to the second and fourth order accurate schemes. This is certainly an exaggeration but
serves as a rough upper bound estimate).
In addition to the cost associated with the update of the child grid, its creation, step {\bf A}, 
via fourth (sixth) order interpolation adds a cost of five (eight) operations
per point. Since we adopt a direct injection (i.e. just copying the child grid value
on the parent's at the corresponding point), the floating point operation 
associated with step {\bf D}.
Hence, the cost of ${\bf B} + {\bf C}$ is $28 N$ and $46 (2N)$. 
Combining these numbers one has a total cost of for the second order
scheme of $T_2 = 23 M + 28 N$ and $T_4 = 38M + 92N$. 

Assuming $M \simeq N$ we have for the $3D$ problem $T_2 \simeq (51 N)^3$ and
$T_4 \simeq (130 N)^3$. Thus, $T_2/T_4 \simeq 3^3$. 
Therefore, the 2nd order update scheme could be employed (in this simple and
conservative estimate) in a grid that is  
$3$ times more refined than that of the fourth order scheme with comparable cost.
This however does not take into account the errors in the solution obtained
with both methods.

To obtain an estimate of this, we can make use of the analysis in \cite{ray,jameson}
where a comparison of the number of points needed to reach a tolerance error per
wavelength is presented.
As can be seen in the quoted numbers in Table \ref{table}, for a target error $< 10^{-2}$
the fourth order accurate scheme is more efficient than the second order one.

We here quote some relevant numbers from~\cite{ray},\\
\begin{table}[ht]
\begin{center}
\begin{tabular}{|c|c|c|c|}
\hline $\epsilon$ & 2nd & 4th & 6th  \\ 
\hline $10^{-2}$ & 26 &  8 & 6 \\ 
\hline $10^{-3}$ & 81 & 15 & 9 \\ 
\hline $10^{-4}$ & 257 & 27 & 13 \\ 
\hline $10^{-5}$ & 816 & 48 & 19 \\
\hline 
\end{tabular} \label{table}
\end{center}
\caption{Number of points required per wavelength to achieve
a target error $\epsilon$ with second, fourth and sixth order
accurate operators.}
\end{table}

Hence, the associated extra-cost of a higher order accurate scheme done via
the tapered grid approach which discards a non-trivial number of points is
offset by the gains in accuracy, and of course, stability of the scheme.

\section{Interpolation Operators}
\label{interpolator}
\subsection{Four point stencils}
Points in the child grid lying (as seen from the parent grid) at $x_{i}$ ($i=1..N$) are directly
injected from the parent value. The points lying in between coarse points at $x_{i+1/2}$ ($i=1..N-1$)
are obtained by the interpolating polynomial,
\begin{eqnarray}
u_{i+1/2} &=& \frac{1}{160} \left ( 9 (u_{i} + u_{i+1}) - (u_{i-1} + u_{i+2})  \right ) \label{int4}
\end{eqnarray}
When excision is employed, the stencil for the second second child grid point must be modified
as there are no enough points to one side of it. In this case we employ,
\begin{eqnarray}
u_{3/2} &=& \frac{1}{128}  \left ( 35 u_1 + 140 u_2 - 70 u_3 + 28 u_4 - 5 u_5\right )
\end{eqnarray}

\subsection{Six point stencils}
As in the previous case, points in the child grid lying (as seen from the parent grid) at $x_{i}$ ($i=1..N$) are directly
injected from the parent value. The points lying in between coarse points at $x_{i+1/2}$ ($i=1..N-1$)
are obtained by the interpolating polynomial,
\begin{eqnarray}
u_{i+1/2} &=& \frac{1}{256} \left ( 150 (u_{i} + u_{i+1}) - 25 (u_{i-1} + u_{i+2}) + 3  (u_{i-2} + u_{i+3}) \right ) 
\end{eqnarray}
When excision is employed, the stencil for the second and fourth child grids point must be modified
as there are no enough points to one side of it. In these cases we employ,
\begin{eqnarray}
u_{3/2} &=& \frac{1}{256} \left ( 63 u_1 + 315 u_2 - 210 u_3 + 126 u_4 - 45 u_5 + 7 u_6 \right ) \\
u_{5/2} &=& \frac{1}{256} \left ( -7 u_1 + 105 u_2 + 210 u_3 - 70 u_4 + 21 u_5 - 3 u_6 \right ) 
\end{eqnarray}

\acknowledgements
This research was supported in part by the NSF under Grants No: PHY0244335,
PHY0326311, INT0204937 to Louisiana State University  and
NSF Grant No. PHY0325224 to Long Island University; CONICET and
SECYT-UNC. Many of these
computations were carried on NPACI computer resources under
allocation awards PHY040021 and PHY040027. L. L. was partially
supported by the Alfred P. Sloan Foundation.

For hospitality while completing parts of this work we wish to thank
the Louisiana State University (O.R \& S.L.), Long Island University (L.L \& O.R.)
and the Isaac Newton Institute for Mathematical Sciences (L.L \& O.R.).

We thank Marsha Berger, Matt Choptuik, Peter Diener, 
Heinz-Otto Kreiss, Jorge Pullin, Frans Pretorius, Erik Schnetter
and Manuel Tiglio  for discussions and/or comments on the manuscript.

\end{document}